\documentclass[
  ]
  {aa}

\usepackage{graphicx}
\graphicspath{{figures/}}
\usepackage[varg]{txfonts}
\usepackage{commath}
\usepackage[super]{nth}

\usepackage{siunitx}
\DeclareSIUnit{\arcsec}{arcsec}
\DeclareSIUnit{\jansky}{Jy}
\DeclareSIUnit{\deg}{deg}

\usepackage[dvipsnames]{xcolor}
\definecolor{click_color}{RGB}{46,48,118}

\usepackage{hyperref}
\hypersetup{
  pdffitwindow=true,
  pdfstartview={FitH},
  pdftitle={Out-of-focus holography at the Effelsberg telescope},
  pdfauthor={Tomas Cassanelli},
  pdfcreator={Tomas Cassanelli},
  pdfproducer={Tomas Cassanelli},
  pdfnewwindow=true,
  colorlinks=true,
  linkcolor=click_color,
  citecolor=click_color,
  urlcolor=click_color
  }

\usepackage{cleveref}

\begin{document}

  \title{Out-of-focus holography at the Effelsberg telescope}
  \subtitle{
    Systematic measurements of the surface of a 100-m telescope using OOF holography
    }

  \author{
    T. Cassanelli\inst{\ref{inst1}}\inst{\ref{inst2}}
    \and U. Bach\inst{\ref{inst1}}
    \and B. Winkel\inst{\ref{inst1}}
    \and A. Kraus\inst{\ref{inst1}}
    }
  \institute{
    Max-Planck-Institut f\"{u}r Radioastronomie, Auf dem H\"{u}gel 69, 53121 Bonn, Germany. \label{inst1} \\
    \email{ubach@mpifr-bonn.mpg.de}
    \and
    Department of Electrical Engineering, Universidad de Chile, Av.\ Tupper 2007, Santiago 8370451, Chile \\
    \email{tcassanelli@ing.uchile.cl} \label{inst2}
    }


  \abstract
  {Out-of-focus (OOF) holography can be used to determine aperture deformations of radio telescopes that lead to errors in the phase of the complex aperture distribution. In contrast to traditional methods, OOF holography can be performed without a reference antenna, which has a number of practical advantages.}
  {The aim of this work is to develop a standard procedure for OOF holography at the Effelsberg telescope. This includes performing OOF holography observations and the development of a dedicated software, the \texttt{pyoof} package, to compute aberrations of the telescope's optical system.}
  {Based on the OOF holography method developed at the Green Bank telescope, we adapted the approach to the Effelsberg 100-m telescope in order to determine the aberrations of the aperture phase distribution (phase-error maps).}
  {The developed OOF holography software is presented as well as the results from observations at Effelsberg. Early results reveal gravitationally-caused residual deformation not contained in the previously existing aperture and pointing model, and hence we propose to make changes to the model to counteract aberrations in the telescope's surface.}
  {The OOF holography method (observations and \texttt{pyoof} package) works as expected at the Effelsberg 100-m telescope and is able to validate the good performance of the existing finite element model. Test measurements show that slight improvements of the aperture efficiency and gain elevation dependence are possible but limited in the current configuration.}


  \maketitle

  \section{Introduction}
  \label{sec:intro}

  The sensitivity (gain) of a single-dish radio telescope, $\Gamma$ \unit{\kelvin\per\jansky}, is a fundamental instrument property that indicates the telescope's response to transform the source's flux density $S_\nu$ into electric potential, or similarly its change in antenna temperature, $T_\text{a}$.
  The sensitivity can be expressed in terms of the aperture efficiency as,
  \begin{equation}
    \Gamma \equiv \frac{T_\text{a}}{S_\nu}= \frac{A_\text{e}}{2k_\text{B}} = \frac{\pi D^2_\text{p}}{8k_\text{B}} \varepsilon_\text{aper}, \quad\text{with}\quad A_\text{e}=A_\text{p}\varepsilon_\text{aper},
    \label{eq:sensitivity}
  \end{equation}
  where $A_\text{e}$ is the effective telescope aperture, which is associated with the aperture efficiency $\varepsilon_\text{aper}$, and the physical area of the primary reflector, $A_\text{p}$. Improving the sensitivity (or increasing $\varepsilon_\text{aper}$) is of imperative scientific importance because: it allows the detection of fainter compact sources, and it reduces the amount of error in the observed power pattern, which directly improves the imaging quality when mapping extended sources.

  The aperture efficiency of a paraboloidal antenna,
  \begin{equation}
    \varepsilon_\text{aper}=\varepsilon_\text{cr}\varepsilon_\text{bk}\varepsilon_\text{ph},
    \label{eq:aperture_efficiency}
  \end{equation}
  is influenced by several factors: the cross-polarization efficiency, $\varepsilon_\text{cr}$; the aperture blockage efficiency, $\varepsilon_\text{bk}$; and the phase-error efficiency, $\varepsilon_\text{ph}$. For well-calibrated apertures, the latter is usually dominated by random surface errors, $\varepsilon_\text{ph}\approx \varepsilon_\text{rs}$. Thus, improving the random-surface-error efficiency\footnote{The random-surface-error efficiency could be gravitational, thermal, and/or other sources of deformations, defined by the empirical law \citep{1966IEEEP..54..633R}, \begin{equation} \varepsilon_\text{rs} = \exp\sbr{-\del{\frac{4\pi}{\lambda}\delta_\text{rms}}^2},\label{eq:e_rs} \end{equation}
  with $\delta_\text{rms}$ the root-mean-square (rms) deviation of the dish in length units. A root-mean-square error $\delta_\text{rms}\ll\lambda$ leads to nearly \qty{100}{\percent} efficiency.
  }, $\varepsilon_\text{rs}$, can increase the sensitivity of a telescope considerably \citep[see][Chapter~7]{stutzman1998antenna}.
  These surface deformations in a single-dish radio telescope have different sources, e.g.: gravitational deformations (mainly due to changes in elevation), thermal deformations (heating or cooling of the mechanical structure), inaccuracies in the panel positions defining the telescope's surface, deformations due to wind, and aging effects. Some of the effects, such as thermal and wind deformations, are extremely hard to model as the influence depends on many parameters, i.e., ambient temperature, illumination by the Sun (and thus its position to the dish), wind strength and direction, etc., which can create differential changes.

  Holography for radio telescopes is a method to measure the main reflector's profile by mapping the amplitude and phase of the entire radiation pattern (\qty[parse-numbers=false]{4\pi}{\steradian}). The mapping is usually done at several elevations with additional equipment, e.g., the use of a smaller reference antenna (near-field, \citealt{2007IAPM...49...24B}), drones equipped with a beacon system, or satellite transit over the primary dish (far- and near-field).

  OOF holography is a phase-retrieval method developed by \citet{2007A&A...465..679N}, which is able to obtain low resolution phase aberrations caused by a telescope's surface inaccuracies and the current setup of the optical system. 
  In contrast to the standard approach, OOF holography has the following advantages: the lack of extra equipment (by observing bright point-like sources), relatively fast measurements ($\sim$45 minutes), and the fact that it offers no restriction in elevation, as previous phase-coherent method applied to Effelsberg \citep{kesteveen2001effelsberg}.
  OOF holography is accomplished by performing several observations of a compact source with a signal-to-noise ratio $\geq200$ \citep[suggested by][]{2007A&A...465..679N}, one of which is in-focus and two of which are out-of-focus (by adding a known axial offset to the sub-reflector), within a short time range to minimize changes in focus, gain, and spanned over a mean elevation (while tracking and mapping the source).

  Large-scale aberrations can be found with OOF holography, and in principle a model can be established when such deformations are stable over longer timescales. The method has successfully been implemented at the Green Bank Telescope (GBT; \citealt{2007A&A...465..685N}), Tianma Radio Telescope \citep{2018ITAP...66.2044D}, and Sardinia Radio Telescope \citep{2020SPIE11445E..6GB}.
  We note that gravitationally-caused deformations are the most well-behaved of the various deformations. Besides this, other errors in the surface e.g., mis-collimations and phase-error on the receiver can also be corrected using this technique. Thermal deformations can be studied (although many more observations are required), but wind and aging effects are discarded due to the short duration and scale of the effects, respectively.

  The OOF holography observations need to be processed by a special software in order to solve the under-determined problem of the power pattern, $P(u, v)$, and aperture distribution, $\underline{E_\text{a}}(x, y)$. This nonlinear numerical problem was approached by developing a modern and specific software for the Effelsberg telescope (which has been now expanded for other single-dish telescopes), the \texttt{pyoof} package (Python). The software requires OOF holography observations as an input and returns the phase-error maps which effectively show surface deformations of the optical system.

  The Effelsberg 100-m single-dish telescope has an active surface control system equipped with \num{96} actuators in its sub-reflector, which has a diameter of \qty{6.5}{\metre}. 
  Currently, the active surface control system is based on a look-up table (with corrections for the sub-reflector provided every ten degrees in elevation) generated from a mechanical structure analysis of the antenna using the finite element method (FEM), i.e., a purely theoretical model \citep[see][Chapter~5]{doyle2002integrated}.
  When the telescope is observing, the needed active surface shape is computed by interpolating between elevation in the look-up table (elevation and actuator displacement FEM model). The current setup does not include real measurements and is solely based on a model of the telescope's mechanical structure.
  An improvement to this approach could be achieved by using the phase-error maps from OOF holography. It is possible to counteract the deformation of the primary dish by using the sub-reflector's active surface with corrections from OOF holography, and hence, increase the telescope's sensitivity.

  In this paper we provide an introduction to OOF holography and its basic theory (\cref{sec:oofh_basics}), and we also introduce a new software that deals with the Effelsberg geometrical aspects and a modern approach to the method (\cref{sec:software}). The final sections are dedicated to the requirements and procedure to perform OOF holography observations at the Effelsberg telescope (\cref{sec:observations}) and the results from these campaigns (\cref{sec:results}), and a discussion of the results and performance of the method are presented in \cref{sec:discussion,sec:conclusions}.
  In addition, detailed information is available in the appendices: Zernike circle polynomials and the convention used in this analysis in \cref{ap:zernike_circle_polynomials}, the geometrical properties of the Effelsberg telescope in \cref{ap:effelsberg_telescope_geometry} and \cref{ap:gregorian_telescope_opd}, and properties of the active surface control system in \cref{ap:effelsberg_active_surface_control_system}.

  The presented work is the continuation of the author's master of science thesis at the Max-Planck-Institut f\"ur Radioastrononmie and Universit\"{a}t Bonn \citep{cassanelli2017oofh}, motivated by the early OOF holography study at the Effelsberg telescope \citep{2014evn..confE..36B}.

  \section{Out-of-focus holography basics}
  \label{sec:oofh_basics}

  The purpose of any holography method is to derive the aperture distribution \citep[see][Chapter~5]{2017isra.book.....T}, $\underline{E_\text{a}}(x, y)$, from an observed map of the telescope beam (antenna power pattern), $P^\text{obs}(u, v)$. These methods can be subdivided into interferometric and non-interferometric. The first corresponds to the with-phase family of methods, which requires the use of a geostationary satellite, reference antenna and/or special receivers (Fresnel field regime; \citealt{2007IAPM...49...24B}). The non-interferometric methods are phase-retrieval algorithms and OOF holography.
  
  The concept of OOF holography is based on the relation between the aperture distribution and phase-error (aperture phase distribution), $\varphi(x, y)$,
  \begin{equation}
    \underline{E_\text{a}}(x, y) = B(x, y)\cdot E_\text{a}(x, y) \cdot \mathrm{e}^{\mathrm i \cbr{\varphi(x, y) + \frac{2\pi}{\lambda}\delta(x,y;d_z)}},
    \label{eq:aperture_distribution}
  \end{equation}
  where $B(x, y)$ corresponds to the telescope's blockage distribution or truncation in the aperture plane, and $E_\text{a}(x, y)$ is the illumination function (or apodization). There is an additional contribution to the phases, which is the optical path difference (OPD), $\delta(x, y;d_z)$. This extra term is necessary for the out-of-focus measurements, when the sub-reflector is slightly moved along the optical axis to defocus the system. The OPD and blockage depend only on the telescope geometry. Unfortunately, it is not possible to measure the aperture distribution directly, but only the power pattern
  \begin{equation}
    P(u, v) = \enVert{\mathcal{F}\sbr{\underline{E_\text{a}}(x, y)}}^2 = \enVert{\iint\limits_{S_\text{a}} \underline{E_\text{a}}(x,y)\cdot\mathrm{e}^{-\mathrm{i}2\pi (xu+yv)}\, \dif S}^2,
    \label{eq:power_aperture_def}
  \end{equation}
  which means that the problem is degenerate (with $\mathcal{F}$ the Fourier transform; FT). The OOF holography method aims to break this degeneracy by utilizing a series of beam maps, one in-focus and two or more out-of-focus, with the latter being performed by adding a axial offset to the sub-reflector of the telescope. This axial offset, $d_z$, is known a priori and it is used to solve the under-determined relation in \cref{eq:power_aperture_def}.

  \subsection{Zernike circle polynomials}
  \label{sec:zernike_circle_poly}

  A convenient approach to solve, \cref{eq:power_aperture_def}, for the aperture phase distribution is to express $\varphi(x, y)$ with a set of orthonormal polynomials and use an optimization procedure to find the best parameters. A good choice are the Zernike circle polynomials, $U^\ell_n(\varrho,\vartheta)$ (expressed in polar basis: $0\leq\varrho\leq1$ and $\qty{0}{\radian}\leq\vartheta\leq\qty[parse-numbers=false]{2\pi}{\radian}$). The Zernike circle polynomials are orthonormal on the unitary circle and their low (radial) orders are closely related to classical optical aberrations; in addition, the lower orders represent large-scale aberrations across the aperture plane.
  The parametrization of the aperture phase distribution is then,
  \begin{equation}
    \varphi(x, y) = \sum_n \sum_\ell K_{n\,\ell} U^\ell_n(\varrho,\vartheta),
    \label{eq:aperture_phase_distribution}
  \end{equation}
  where the set of coefficients, $K_{n\,\ell}$, needs to be found. The constants $n$ and $\ell$ are the order (or degree) and angular dependence of the polynomials.

  The classical way to represent the Zernike circle polynomials is by the radial polynomials \citep{1965poet.book.....B},
  \begin{equation}
    R^{\pm m}_n(\varrho)  = \frac{1}{\intoo{\frac{n-m}{2}} \hspace{-3pt}  \raisebox{-.7ex}{\scalebox{1.5}{!}} \, \varrho^m} \cbr{\frac{\dif}{\dif \,(\varrho^2)}}^{\frac{n-m}{2}}\cbr{\intoo{\varrho^2}^{\frac{n+m}{2}}\intoo{\varrho^2-1}^{\frac{n-m}{2}}},
    \label{eqd:radial}
  \end{equation}
  with $m=|\ell|$; the order, $n$, a positive integer number; and $\ell$, the angular dependence, which can be a positive or negative integer. For each angular dependence the order is restricted as $n=|\ell|, |\ell| + 2, |\ell|+ 4, \dotso$. The Zernike circle polynomials are then given by
  \begin{equation}
    U ^{\ell}_n(\varrho,\vartheta)  =  \begin{cases} R^m_n(\varrho)\cos m\vartheta & \ell \geqslant0 \\ R^m_n(\varrho)\sin m\vartheta & \ell<0 \end{cases}.
    \label{eq:zernike}
  \end{equation}
  There are several conventions on how to sort the polynomials with their constants $K_{n\,\ell}$. In \cref{ap:zernike_circle_polynomials} we show the current convention used in the presented analysis and \texttt{pyoof} software (\cref{sec:software}).

  \subsection{Power pattern fit}

  The optimization procedure tries to minimize the residuals
  \begin{equation}
    \min \left\{P^\text{obs}_\text{norm}(u, v) - P^\text{model}_\text{norm}(u, v)\right\},
    \label{eq:min_residual}
  \end{equation}
  of the observed power patterns ($P^\text{obs}_\text{norm}$) relative to the model patterns ($P^\text{model}_\text{norm}$) where the subscript norm represents the normalized power pattern,
  \begin{equation}
    P_\text{norm}(u, v) = \frac{P(u, v) - P_\text{min}}{P_\text{max} - P_\text{min}}.
  \end{equation}
  The minimization is technically performed by means of a nonlinear least-squares algorithm. The model power patterns can be computed from the model aperture distribution, \cref{eq:aperture_distribution}, via FT, \cref{eq:power_aperture_def}. Since there are three or more observed beam maps, a simultaneous fit ought to be done, which yields the Zernike circle polynomial coefficients, $K_{n\,\ell}$, representing $\varphi(x, y)$. 
  One important detail is that in order to use a fast FT (FFT) algorithm, which greatly speeds up the process, a proper spatial grid must be chosen---ideally one could create a grid for the model that is matching the pixel grid of observations---but it is also possible to use a different spatial grid for the aperture distribution and then simply re-grid the resulting model power pattern to the observed grid.

  \subsection{Surface improvement}
  \label{sec:surface_improvements}

  Based on the derived aperture phase distribution functions it is possible to apply improvements to the aperture, e.g., by adjusting the panels of the main dish. However, many effects that act on the surface of the dish will be elevation-dependent. For the 100-m telescope at Effelsberg, one has the opportunity to actively readjust the surface of the sub-reflector to correct inaccuracies of the main dish. Since elevation is an important parameter for gravitational induced surface deformation, the OOF holography observations should be performed for a number of different elevations. If a sufficiently large number of phase-error maps is available, it is even possible to create a model of the gravitational changes to the primary reflector, which could eventually replace the existing FEM model that is implemented in the Effelsberg control system and steers the active surface of the sub-reflector. 
  Compared to the FEM model the advantage of the OOF holography method is that the aberrations are derived empirically and can reveal residual errors that are not covered by the theoretical model in the current look-up table. Technically, the displacement of the actuators is defined in physical length, while the aperture phase distribution is in units of wavelengths. The relation between them is
  \begin{equation}
    \varphi_\bot(x, y) = \frac{\lambda}{4\pi} \varphi(x, y),
    \label{eq:displacement}
  \end{equation}
  where $\varphi_\bot(x, y)$ is the displacement distribution for the actuators. The extra factor of ${1}/{2}$ in \cref{eq:displacement} is due to the secondary-focus observations. 
  The ray of light is affected by offsets in the sub-reflector twice: first when it travels to the sub-reflector, and second when it moves from the secondary dish to the secondary focus. We note that \cref{eq:displacement} is only a first-order approximation\footnote{Modelling the actuator distribution would require to know the exact inclination of the feed receiver, and include it in an optical ray tracing software, e.g., ZEMAX.} because it assumes that the rays hitting the sub-reflector surface are always reflected parallel to the $z_f$-axis (see the right-hand side of \cref{fig:effelsberg_geometry}). 

  Another important consideration when deriving the aperture phase distribution evaluation is to completely remove the tilt terms ($x$- and $y$-tilt) and piston from the phase-error. The tilt terms are associated with average telescope movement in the $x$ and $y$ directions (of the phase-error plane, $\varphi(x, y)$). To remove this effect from the Zernike circle polynomials, the piston and tilt are simply set to zero ($K_{0\,0}=K_{1\,1} = K_{1\,-1}=0$). Aberrations and their relation to the Zernike circle polynomials can be seen in \cref{tab:zernike_aberration}.

  \section{Software} 
  \label{sec:software}

  To accomplish all tasks and find the corrections for the aperture phase distribution, the \texttt{pyoof} package\footnote{GitHub \texttt{pyoof} package repository: \url{https://github.com/tcassanelli/pyoof} and \texttt{pyoof} package documentation: \url{https://pyoof.readthedocs.io/en/latest/}.} for the Python programming language was created. The \texttt{pyoof} package makes use of Astropy \citep{2018AJ....156..123A} for physical units support. At the moment, the package is shipped with functions reflecting the Effelsberg 100-m telescope geometry, but it can easily be extended to support other geometries and  configurations, as seen with the preliminary work at the Sardinia telescope \citep{2020SPIE11445E..6GB}.
  The software is separated into four main modules: Zernike circle polynomials (\cref{sec:zernike_module}), the telescope geometry (\cref{sec:telescope_geometry_module}), the aperture-like functions (\cref{sec:aperture_module}), and actuators (Effelsberg-specific only; \cref{sec:actuator_module}).

  \subsection{Zernike module}
  \label{sec:zernike_module}

  The Zernike module provides functions to compute Zernike circle polynomials as defined in \cref{eqd:radial,eq:zernike}. The number of polynomials up to order (or degree) $n$ is $\frac{1}{2}(n+1)(n+2)$. In the literature there is no clear convention regarding the ordering of the polynomials. The \texttt{pyoof} package returns them in the following order: $U^0_0, U^{-1}_1, U^1_1, \dotso, U^{-\ell}_n, \dotso, U^\ell_n$. The convention can be seen in \cref{ap:zernike_circle_polynomials}.

  The orthonormality of the Zernike circle polynomials is only given if the radial polynomials are normalized in \cref{eqd:radial} or \cref{eq:zernike}. However, for our application, orthonormality is not needed and instead the observed and model power patterns are normalized, \cref{eq:min_residual}.

  \subsection{Telescope geometry module}
  \label{sec:telescope_geometry_module}

  The telescope geometry module offers functions to specify the geometrical properties of the telescope, consisting mainly of a description of the blockage distribution and the OPD function.

  \subsubsection{Blockage distribution}

\begin{figure}
    \centering
    \includegraphics{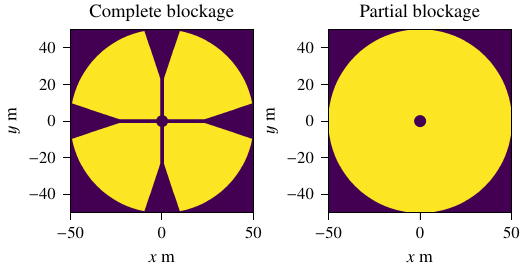}
    \caption{Blockage distribution, $B(x, y)$, Effelsberg telescope. Three effects can be seen (left-hand side): the center region is blocked by the sub-reflector, the cross-like lines originate from the four support struts, as do the triangle-shaped areas in the outskirts. The difference between the two support-leg induced effects is whether a ray is first reflected from the dish and is then (eventually) blocked when it travels from the main dish to the secondary mirror (sub-reflector), or if it is blocked before it can even hit the main dish. The right-hand plot corresponds to the truncation of the sub-reflector only.}
    \label{fig:blockage}
  \end{figure}

  The blockage distribution resembles the elements in the telescope structure that hinder radiation from entering the receiver feed; it is characterized by the reflector configuration. The Effelsberg telescope has a Gregorian setup and its blockage is determined by the primary focus cabin's support struts and the cabin itself (same as in \cref{ap:effelsberg_telescope_geometry} \cref{fig:effelsberg_geometry} right-hand side). Note that the support struts also cause a shadow effect, which appears as a widening of the blocked area towards the outskirts of the aperture. This is because the inner parts of the support struts are closer to the focus cabin. This effect has been describe by several authors, e.g., \citet{1996A&AS..119..115H,kesteveen2001effelsberg}. Examples of a blockage distribution are presented in \cref{fig:blockage}.
  In \texttt{pyoof}, the blockage distribution is represented by a binary array; the blocked elements are zero-valued. The total blockage of the Effelsberg telescope is about \qty{\sim16}{\percent} that results in an efficiency loss ($1-\varepsilon_\text{bk}$) of \qty{\sim30}{\percent}. 
  A correct determination of the blockage shape produces noticeable results in the beam model, i.e., a closer shape to the real (observed) beam.

  \subsubsection{Optical path difference function}
  \label{sec:opd_function}

\begin{figure}
    \centering
    \includegraphics{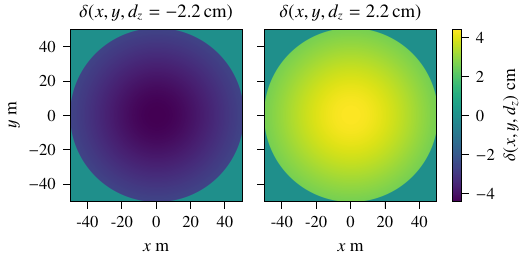}
    \caption{OPD functions as computed in the \texttt{pyoof} package. The left-hand side corresponds to a path difference made with a negative axial offset of $d_z=\qty{-2.2}{\cm}$. The right-hand side axial offset is for a positive offset of $d_z=\qty{2.2}{\cm}$. The OPD is computed analytically in \cref{ap:gregorian_telescope_opd}.}
    \label{fig:opd}
  \end{figure}

  The second geometrical effect is owing to the OPD function. The path difference is related to the reflector configuration in the telescope. Such a relation can be derived by a ray tracing software or computed analytically. For an ideal Cassegrain/Gregorian system the path difference is given by
  \begin{gather}
    \delta(x,y;d_z) = d_z\del{ \frac{1-\mathcal{A}^2}{1+\mathcal{A}^2} + \frac{1-\mathcal{B}^2}{1+\mathcal{B}^2}}, \label{eq:opd} \\
    \text{with} \quad \mathcal{A} = \frac{r'}{2F_\text{p}} \quad\text{and}\quad \mathcal{B}=\frac{r'}{F_\text{eff}}.
  \end{gather}
  The OPD function, $\delta(x, y;d_z)$, depends only on the focal length of the paraboloid, $F_\text{p}$, the effective focal length, $F_\text{eff}$, and the distance from the center of the aperture plane to any $(x, y)$ point in the surface, $r'=\sqrt{x^2 + y^2}$ (see \cref{ap:effelsberg_telescope_geometry}). For a Gregorian configuration a detailed derivation is in \cref{ap:gregorian_telescope_opd}.
  The OPD function effectively acts as an attenuation (apodization) across the aperture plane; see \cref{fig:opd}.

  \subsection{Aperture module}
  \label{sec:aperture_module}

  The aperture module allows for the construction of the aperture distribution from the various constituents. In addition to the blockage, the illumination function and the aperture phase distribution are needed.

  \subsubsection{Illumination function}
  
  Receivers in a radio telescope utilize antenna feeds to convert the electromagnetic field into electric signals. The feeds have an antenna pattern themselves, which effectively leads to a weighting of the aperture, also referred to as tapering or apodization. Owing to the large variety of physical feeds (from simple dipoles and feed horns to modern patch antennas used for focal plane arrays), we know of many different (simplifying) functions for the illumination. In the literature, most of them make use of two, three, or up to four parameters. In \texttt{pyoof} the parabolic taper on a pedestal function \citep[see][Chapter~7]{stutzman1998antenna}
  \begin{equation}
    E_\text{a}(x, y) = A_{E_\text{a}}\cbr{C + \del{1-C} \sbr{1 - \del{\frac{R_\text{a}}{\frac{D_\text{p}}{2}}}^2 }^q },
    \label{eq:illumination}
  \end{equation}
  is provided, where $R_\text{a}=\sqrt{(x-x_0)^2+(y-y_0)^2}$ is the offset of the illumination function with respect to the center of the aperture plane; $C=10^{\frac{c_\text{dB}}{20}}$ denotes the so-called taper strength of the illumination; and $q$ another degree of freedom\footnote{For reflector antennas $1\leq q \leq 2$ \citep[see][Chapter~15]{lo2013antenna}.}. The values of $C$ and $q$ should be chosen such that they provide similar edge tapers and amplitude slopes at the edge as compared with the actual feed/reflector configuration.
  The \texttt{pyoof} package also allows one to work with user-defined illumination functions.

\begin{figure}
    \centering
    \includegraphics{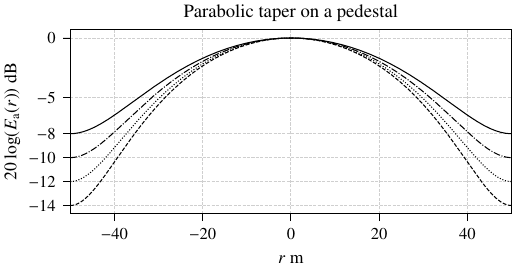}
    \caption{Illumination function of the parabolic taper on a pedestal; see \cref{eq:illumination}. The four curves are centered on the origin with same amplitude ($A_{E_\text{a}}=1$) and $q=2$. From top to bottom the taper values are \qtylist{-8;-10;-12;-14}{\decibel}.}
    \label{fig:illumination}
  \end{figure}

  \Cref{fig:illumination} shows the parabolic taper on a pedestal for four different taper values. The purpose of the illumination is to try to imitate the behavior that the receiver has over the parabolic surface, by setting free parameters $\sbr{A_{E_\text{a}}, c_\text{dB}, q , x_0, y_0}$. 
  In general, the shape of an illumination function (strong in the center, weak at the edges), comes from the need to eradicate spill-over, as well as to minimize the effect of the side lobes over the final power pattern.
  The illumination function has a similar effect on the aperture distribution as the OPD function, with the difference that the OPD function is known a priori (known $d_z$), and the illumination parameters are not (in particular the taper strength $c_\text{dB}$ and free parameter $q$). Nevertheless, this information can be found in the receiver specifications or indirectly by trial and error using the OOF holography method.

  \subsubsection{Aperture phase distribution}
  The aperture phase distribution, $\varphi(x, y)$, is the final product of the \texttt{pyoof} package. In order to find it, a numerical optimization routine is applied to calculate the coefficients, $K_{n\,\ell}$, that minimize the $\chi^2$ residual of the observed and model beam pattern; see \cref{eq:min_residual}. As the problem is nonlinear, \texttt{pyoof} utilizes a (nonlinear) least-squares algorithm, the trust region reflective\footnote{The Python function used for this task is: \url{https://docs.scipy.org/doc/scipy/reference/generated/scipy.optimize.least_squares.html}} (TRR) algorithm \citep[see][Chapter~4]{nocedal2006numerical}, which is a more advanced variant of the well known Levenberg–Marquardt algorithm used by \citet{2007A&A...465..679N,2007A&A...465..685N}. In practice both algorithms yield similar results, but the TRR allows the user to define boundary limits to the coefficients. Furthermore it is slightly faster (in the Effelsberg case). 
  The maximum order of the Zernike polynomials that is fitted is a hyperparameter and must be chosen by the user. For Effelsberg, we find reasonable results in the range between $5\leq n \leq6$. A more detailed discussion is included in \cref{sec:discussion}. 

  \subsubsection{Aperture distribution and power pattern}
  During the optimization, it is necessary to calculate the antenna power pattern to compare it with the antenna observable. The aperture module provides the necessary functions to do that, based on the aperture phase distribution. The power pattern is computed by means of an FFT from the aperture distribution, $\underline{E_\text{a}}$.

  \subsection{Actuator module}
  \label{sec:actuator_module}

  The actuator module deals with the polynomial representation of the phase-error maps in the space of the active surface control system. Actuators are in general a system that will change from telescope to telescope and different geometries, i.e., this is only supported for the Effelsberg telescope at the moment. The actuator module is key while trying to implement the improvements described in \cref{sec:surface_improvements}, and it is an extension to the software rather than a fundamental piece of code. Most OOF holography studies are independent of this module.

  \subsection{Model parameters}
\begin{figure}
    \centering
    \includegraphics{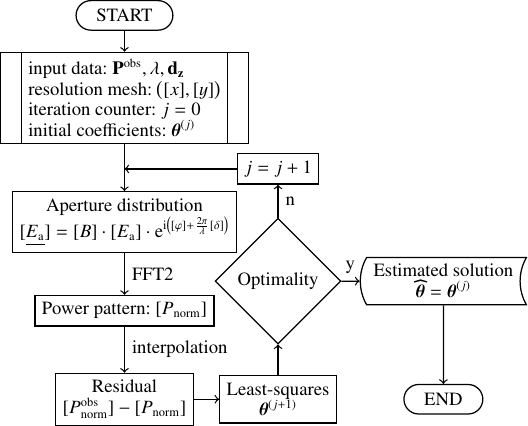}
    \caption{Flowchart and basic structure of the \texttt{pyoof} package. Bold letters and letters in brackets represent data vectors and data meshed arrays, respectively. The set of coefficients to be found (illumination function and Zernike circle polynomials) is given by $\boldsymbol\theta$ (eq.~\ref{eq:coeff_vector}), and $j$ represents the iteration counter.}
    \label{fig:flowchart_pyoof}
  \end{figure}

  The number of parameters to specify the aperture is large. As some of them are well determined by the telescope geometry and other a priori known circumstances, it is advised to keep these fixed. This applies to: the telescope dimensions, blockage distribution, the axial offsets ($d_z$), and the observation wavelength ($\lambda$). The free parameters include: the Zernike circle polynomial coefficients (which describe the aperture phase distribution), the illumination-function coefficients, normalization factor, etc. 
  Internally, these parameters are put into a vector,  
  \begin{equation}
    \boldsymbol{\theta} = \sbr{A_{E_\text{a}}, c_\text{dB}, q, x_0, y_0, K_{0\,0}, K_{1\,-1}, K_{1\,1}, K_{2\,-2}, \dotso, K_{n\,\ell} }^\intercal.
    \label{eq:coeff_vector}
  \end{equation}
  The parameter vector $\boldsymbol\theta$ is then changed in every iteration of the minimization process until the TRR objective is achieved. A basic flowchart for the optimization procedure is visualized in \cref{fig:flowchart_pyoof}. In addition, the implemented version of the TRR provides bounds functionality that can be included if desired, e.g., by limiting $c_\text{dB}\in\sbr{\qty{-21}{\decibel}, \qty{-1}{\decibel}}$. 

  \subsection{Simulations and software test}
  \label{sec:simulations}
\begin{figure*}
    \centering 
    \includegraphics{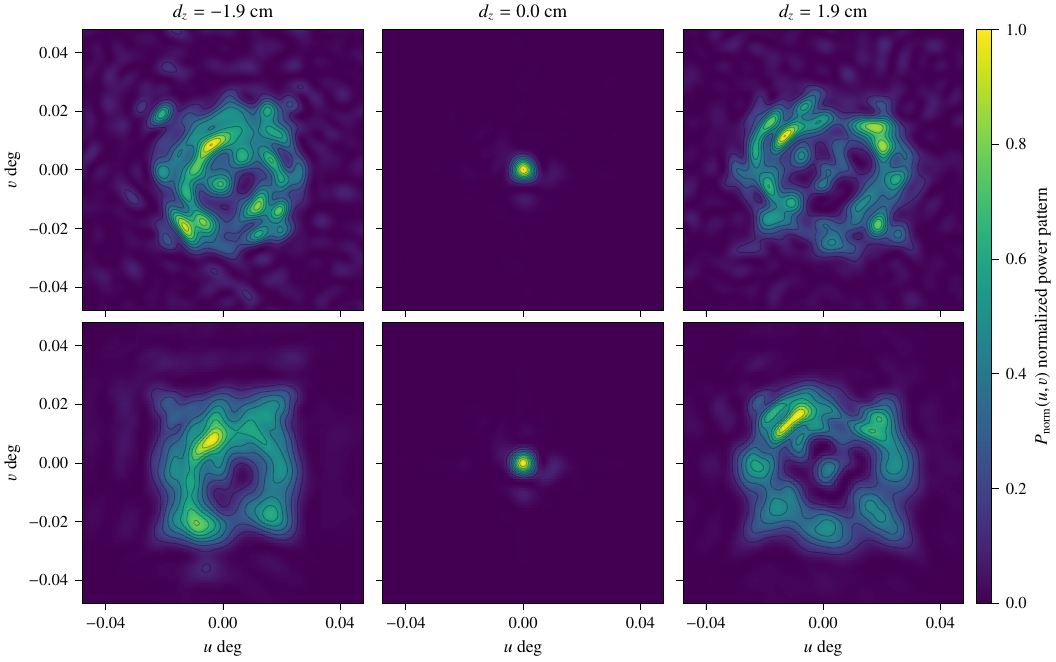}
    \caption{Simulated and computed power patterns. The upper row corresponds to a simulated beam, using up to order $n=5$ for the Zernike circle polynomials, with a centered illumination with $c_\text{dB}=\qty{-14.5}{\decibel}$ taper strength, $q=1.4$, and $(x_0,y_0)=(0, 0)$. The tilt parameters were also set to zero ($K_{1\,1}=K_{1\,-1}=0$). The simulated axial offset is $d_z=2.2\lambda$.
    The left and right columns show the defocused pattern and the center column is the in-focus beam.
    The Zernike circle polynomial coefficients, $K_{n\,\ell}$, were generated randomly with a normal distribution. Gaussian noise was added until a signal-to-noise ratio of \num{750} in the in-focus beam (top middle panel) was achieved.}
    \label{fig:beam_sim_com}
  \end{figure*}

  To test the software simulations were performed. An artificial power pattern was generated based on the Effelsberg geometrical properties (see \cref{sec:observations}) in the following manner.
  An initial set of Zernike circle polynomials coefficients is randomly generated, as well as illumination coefficients, up to an order $n = 5$ ($21 + 5$ dimension for $\boldsymbol\theta$)\footnote{These set of coefficients ($K_{n\,\ell}$) have been chosen randomly (each of them) from a normal distribution with mean $\mu$ and variance $\sigma^2$, which is how the Zernike circle polynomial coefficients and parameters vary at the Effelsberg telescope.}. This yields a beam pattern map for each offset ($d_z = \qtylist[list-pair-separator = {, }]{0;\pm1.9}{\cm}$, at a frequency of $\qty{34.75}{\giga\hertz}$) and we mimic an observation by adding (Gaussian) noise (upper row in \cref{fig:beam_sim_com}). Noise is added until a signal-to-noise ratio of \num{750} is reached in the in-focus beam\footnote{According to \citet{2007A&A...465..679N}, the OOF holography method works for signal-to-noise ratios of at least about 200.} ($d_z=\qty{0}{\centi\m}$).

  Now, the optimization procedure of the \texttt{pyoof} package is applied to derive the best-fit parameters. An initial estimate of \cref{eq:coeff_vector} is required for the least-squares minimization. For the Effelsberg telescope it has been seen that values are within $K_{n\,\ell} \in [-0.3, 0.3]$. The start point will get closer to those values no mater the initial point, but given a good start, the overall process will be faster. Based on previous observations we fixed $c_\text{dB}=\qty{14.5}{\decibel}$, $q=1.4$, and $(x_0, y_0)= (\qty{0}{\cm}, \qty{0}{\cm})$ for the illumination function, and piston and tilt set to zero. Then we have 18 non-trivial Zernike circle polynomials coefficients to fit plus the illumination amplitude. 

\begin{figure*}
    \centering
    \includegraphics{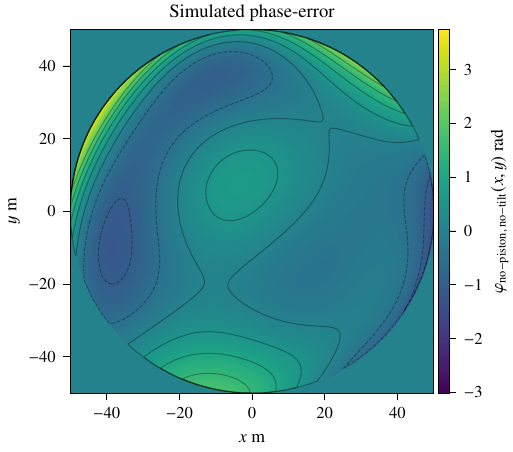}
    \includegraphics{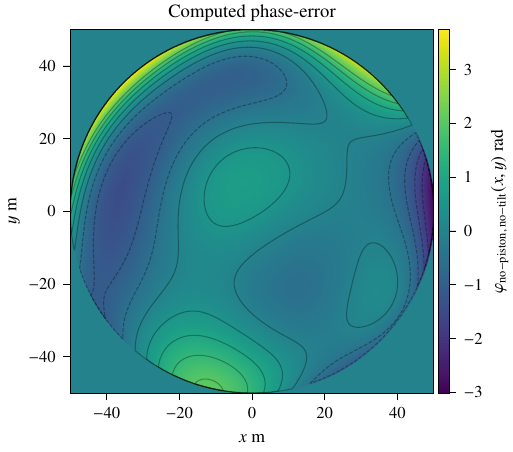}
    \caption{Aperture phase distribution (phase-error maps) of the simulated (left panel) and retrieved (right panel) model. The simulated Zernike circle polynomial coefficients ($K_{n\, \ell}$) were selected randomly from a normal distribution, mean centered ($\mu=0$), and $\sigma$ added until a defocused map reached a signal-to-noise ratio level of \num{750}. The contour lines are between $-2$ to $2$ half-radian intervals.}
    \label{fig:phase_sim_com}
  \end{figure*}

  Based on this, the best-fit power pattern can be calculated (\cref{fig:beam_sim_com} bottom row). Note that the maximum amplitude in the beam maps is close to be \num{1} due to the normalization that was applied. One can also compare the simulated and retrieved aperture phase distributions; see \cref{fig:phase_sim_com}. 
  There are some minor differences between the simulated and retrieved phases, but the (local) minima and maxima have comparable amplitude values and the overall large-scale ($\sim$\qtyrange{10}{20}{\metre}) shape also matches well. It is important to notice that the edges of the aperture phase distribution, in \cref{fig:phase_sim_com}, are not that well constrained and will always have an overall worse estimation. This is mainly due to the illumination function edges.

  \section{Observations} \label{sec:observations}
  To perform OOF holography observations, a moderately high signal-to-noise ratio of a point-like source is required, and it is observed with three beam maps (two out-of-focus and one in-focus). The observations are performed in continuum mode using the so called \textit{on-the-fly} mapping. In this mode the sky is scanned continuously in stripes along the azimuth or elevation direction.

  A compromise between resolution and signal-to-noise ratio has to be found for the observing wavelength. The wavelength will determine the degree of accuracy to which the aperture phase distribution can be measured (see eq.~\ref{eq:displacement}), e.g., using a \qty{9}{\mm} receiver (at \qty{32}{\giga\hertz}), one radian of phase corresponds to \qty{745}{\micro\metre}; and with a \qty{7}{\mm} receiver (at \qty{34.75}{\giga\hertz}), it is \qty{686}{\micro\metre}. On the other hand, a longer-wavelength receiver will obtain a higher signal-to-noise ratio as many radio sources have steep spectra and receivers have lower system temperatures at longer wavelength. A higher signal-to-noise ratio will allow the fitting of higher order polynomials, which will determine the possible spatial resolution that can be obtained.
  The observations in this study were performed with the Effelsberg \qty{7}{\mm} and \qty{9}{\mm} receivers. Both are wide-band receivers with good sensitivity. The technical details can be found in \cref{tab:receivers}. At frequencies of \qtyrange{30}{38}{\GHz} the atmospheric conditions do not depend so much on the weather and therefore this frequency range was chosen. The first observations were conducted with the \qty{9}{\mm} receiver, because the newer \qty{7}{\mm} receiver was not available at that time. Later the \qty{7}{\mm} was used, because of backend problems at the \qty{9}{\mm} receiver. \Cref{tab:obs_config} shows some basic parameters of the beam maps and OOF holography observations. The size of the defocused beam pattern determines the size of the map, so that the full beam pattern is covered plus some overhead to fit the baseline. The spacial separation between the sub-scans (stripes of the map) has to be lower than half of the beam width of the antenna to properly sample the map. 

  \begin{table}[t]
    \centering
    \caption{Technical properties of the receivers used for the OOF holography observations. Given are the name, the possible frequency range, center frequency used, detected bandwidth (BW), system temperature ($T_\text{sys}$), system equivalent flux density (SEFD; $S_\text{sys}$), and the full with half maximum of the beam (FWHM).}
    \label{tab:receivers}
    \begin{tabular}{lcccrrr}
      \hline
      \hline
      Name & Freq. & Freq. & BW & $T_\text{sys}$ & SEFD & FWHM \\
        & range & used & & & & \\
        & \unit{\giga\hertz} & \unit{\giga\hertz} & \unit{\giga\hertz} & \unit{\kelvin} & \unit{\jansky} & \unit{\arcsec} \\
      \hline
      \qty{9}{\mm} & \numrange{30}{34} & \num{32.00} & \num{4.0} &  \num{68 }& \num{85} & \num{24.5}\\
      \qty{7}{\mm} & \numrange{33}{50} & \num{34.75} & \num{2.5} &  \num{100} & \num{160} & \num{23.0}\\
      \hline
    \end{tabular}
  \end{table}

  Before the beam images can be analyzed by the OOF holography software, the raw \texttt{FITS} data from the antenna has to be pre-processed. The basic reduction includes clipping and gridding of the on-the-fly scans to square pixels, base level adjustment, and the conversion to antenna temperatures using the known temperature of a noise source. For these steps the Effelsberg continuum data reduction software toolbox and \texttt{NOD3} is used \citep{2017A&A...606A..41M}. Opacity corrections were calculated from measurements of the water vapor radiometer at Effelsberg and the final calibration to Jansky was obtained from observations of primary calibrators. \Cref{tab:sources} list the primary calibrator sources and the flux densities of the sources used for OOF holography observations, which have to be bright enough to reach the required signal-to-noise ratio. To further improve the signal-to-noise ratio, maps observed in different scanning directions (azimuth and elevation) can be combined with the basket-weaving method \citep{1988A&A...190..353E}. All three maps are exported from the \texttt{NOD3} software together into a single \texttt{FITS} file for the final analysis in the \texttt{pyoof} package.

  \begin{table}[t]
    \centering
    \caption{Properties of the OOF holography observations. Given are the receiver name, central frequency, maps size, scan speed, amount of defocus for the OOF holography maps, and the obtainable resolution for $\varphi = \qty{1}{\radian}$ of phase-error.}
    \label{tab:obs_config}
    \begin{tabular}{lllcll}
      \hline
      \hline
      Name & Freq. & Maps size & Scan speed & $d_z$ & $\varphi_\bot$ \\
        & \unit{\giga\hertz} & \unit{\arcsec\squared} & \unit{\arcsec\per\second} & \unit{\cm} & \unit{\micro\metre} \\
      \hline
      \qty{9}{\mm} & \num{32.00} & $360\times360$ & \num{40.0} & $\pm2.2$ & \num{745}\\
      \qty{7}{\mm} & \num{34.75} & $340\times340$ & \num{22.8} & $\pm1.9$ & \num{686}\\
      \hline
    \end{tabular}
  \end{table}

  Each OOF holography observation set takes about 45 minutes to complete and is comprised of one in-focus beam map and two OOF holography beam maps. Each cycle of three maps starts with a pointing and focus scan to center the source and optimize the focus. After the first map at perfect focus the two defocused maps with axial offset of $\pm d_z$ are measured. Prior to each map the antenna pointing is checked to properly align all three images.
  
  While tracking the target source over these $45$ minutes, the telescope will experience changes in azimuth and elevation. These changes will depend on source and observing time and could be up to $\delta\alpha \lesssim\qty{10}{\deg}$ from the start to end of a single OOF holography observation (a single phase-error map). The elevation estimate for a phase-error map $\varphi^{(\alpha)}(x, y)$ is the mean elevation over the entire OOF holography observation (three beam map scans).

  Processing of the observations was carried out in a similar manner to that described in \cref{sec:simulations}, where piston and four illumination parameters were fixed (all except the illumination amplitude), but tilt terms were added to account for pointing errors in the observation. This is ${20 + 1}$ parameters to fit with \texttt{pyoof}.

  \begin{table}[t]
    \centering
    \caption{Radio sources used for the OOF holography imaging and calibration of the data. Given are the source name, right ascension and declination coordinates in J2000 (RA/Dec), flux density ($S_\nu$) at \qty{34}{\giga\hertz} ($S_{\qty{34}{\giga\hertz}}$), and a qualifier (Q) for target (T) and calibrator (C). The flux densities of the target sources are typical flux densities from 2020. They are variable active galactic nuclei (AGN), but are usually bright and the long term variability does not affect the OOF holography observations.}
    \label{tab:sources}
    \begin{tabular}{lccrc}
      \hline
      \hline
      Name & RA & Dec & $S_{\qty{34}{\giga\hertz}}$ & Q \\
      \hline
      3C286   & 13h31m08.3s & \ang[parse-numbers=false]{+30;30;03.0} & \qty{1.72}{\jansky} & C\\
      NGC7027 & 21h07m01.6s & \ang[parse-numbers=false]{+42;14;10.0} & \qty{5.25}{\jansky} & C\\
      3C84    & 03h19m48.2s & \ang[parse-numbers=false]{+41;30;42.1} & \qty{21.80}{\jansky} & T\\
      3C273   & 12h29m06.7s & \ang[parse-numbers=false]{+02;03;08.6} & \qty{11.50}{\jansky} & T\\
      3C454.3 & 22h53m57.8s & \ang[parse-numbers=false]{+16;08;53.6} & \qty{11.30}{\jansky} & T\\
      \hline
    \end{tabular}
  \end{table}

  \subsection{August 2017 observation campaign} \label{sec:ObsAug2017}

\begin{figure}[t]
    \centering
    \includegraphics{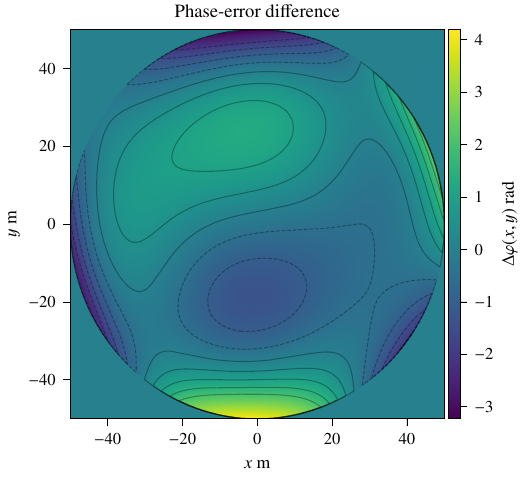}
    \caption{Phase-error orientation and the active surface control system. We observe 3C84 with active surface turned off at a high elevation (where deformations are stronger) and subtracted another phase-error map while observing at a similar elevation with active surface turned on, $\Delta\varphi = \varphi^{(\qty{76.05}{\deg})} - \varphi^{(\qty{69.6}{\deg})}$. The result yields a clear and similar pattern to that seen in \cref{fig:fem_lookup}. The contour lines are between -2 to 2 half-radian intervals.}
    \label{fig:phase_convention_9mm}
  \end{figure}

  This set of observations corresponds to the first and earlier tests for the \texttt{pyoof} package and its development. To better understand the system, and specifically the conventions used (e.g., coordinate systems, ordering of Zernike circle polynomial coefficients, labeling of the actuators in sub-reflector, and others) observations were taken to study the orientation for the aperture phase distribution, $\varphi^{(\alpha)}(x, y)$, maps and the actuators on the active surface control system. Several tests were performed with the \qty{9}{\milli\meter} receiver; the most prominent of them were observations at high elevation, where the main dish requires more corrections (a larger actuator amplitude, see \cref{ap:effelsberg_active_surface_control_system}) in its surface. This behavior of the primary dish has also been seen by \cite{2015JAGeo...9....1H}, where the correction at a high elevation oscillated between $\qty{\pm5}{\milli\meter}$.
  One of these tests was performed at a mean elevation of $\alpha=\qty{76.05}{\deg}$, for 3C84 (only with a scan across azimuth), and with the active surface turned \textit{off}, and another at mean elevation $\alpha=\qty{69.6}{\deg}$ with active surface \textit{on}. Then both observations were processed with \texttt{pyoof} and the difference ($\Delta\varphi(x, y)$) was calculated; as before, tilt and piston terms were neglected. The result in \cref{fig:phase_convention_9mm} shows the classical aberration pattern from the dominant first and second coma (see \cref{tab:zernike_aberration} and \cref{fig:zernike_convention}). As can be seen in the figure, this is similar to the theoretical FEM model in \cref{fig:fem_lookup} (\cref{ap:effelsberg_active_surface_control_system}).

  \subsection{December 2019 observation campaign} \label{sec:ObsDec2019}

\begin{figure*}
    \centering
    \includegraphics{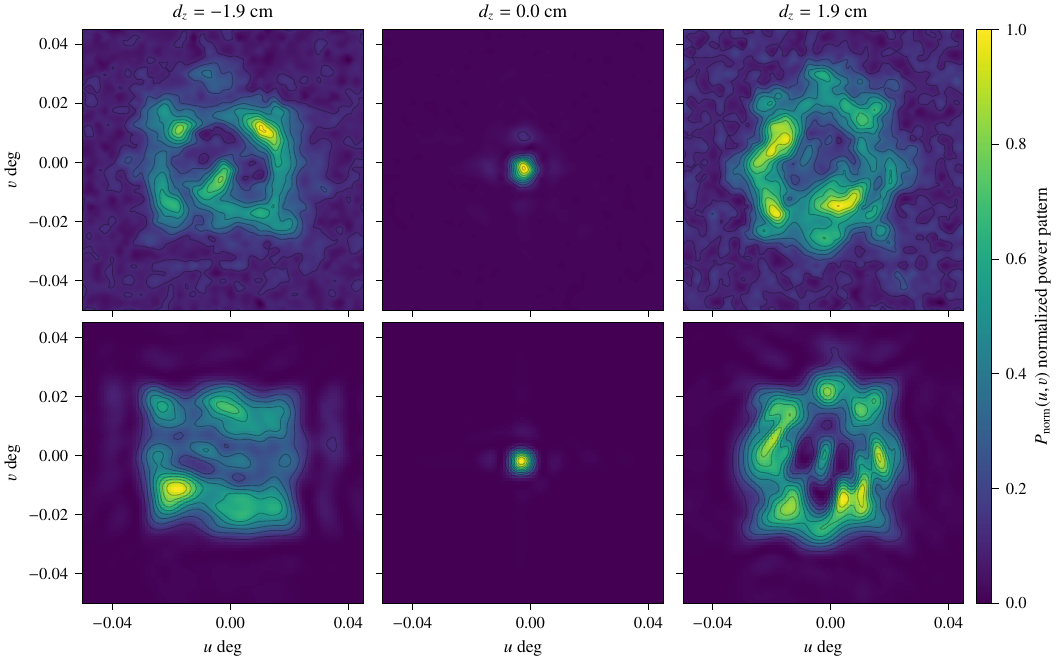}
    \caption{Observed (December 2019) and computed power patterns. The upper row corresponds to the 3C84 observed beam, using the \qty{7}{\mm} receiver at a mean elevation of \qty{48}{\deg} (azimuth scans). The lower row shows the best fit for an $n=5$. The observed focused beam (top centered panel) has a signal-to-noise ratio of \num{610}. Notice that in the observing case the signal-to-noise ratio of the $\pm d_z$ maps may be noisier than the simulated case in \cref{fig:beam_sim_com}. Observing with a defocused telescope degrades further the imaging, increasing the observed noise level.}
    \label{fig:beam_obs_com}
  \end{figure*}

  For these two days of observation (December \nth{4} and \nth{5}, \num{15} observation hours), the source 3C84 was mapped; calibration was done by cross-scans in azimuth and elevation on 3C286 and NGC7027. Ten OOF holography beam maps were taken from \qtyrange{34}{78}{\deg} elevation range. An example of these observations can be seen in the upper row of \cref{fig:beam_obs_com}.

  The observations' maps size was $\qty{342}{\arcsec}\times\qty{342}{\arcsec}$ with a beam size of \qty{22}{\arcsec} using the \qty{7}{\mm} receiver. In this configuration the scan speed is \qty{22.8}{\arcsec\per\second} and the sub-scans for each row are  separated by \qty{9}{\arcsec} (within a single beam map).

  Before every set of three maps (with focus offsets of: \qtylist{0;+1.9;-1.9}{\cm}) the optimal focus along the $z$-axis (see \cref{fig:effelsberg_geometry}) was measured and before every map the antenna was pointed on the target source with a cross-scan. The scan direction of each set was alternated between azimuth (\textit{L}) and elevation (\textit{B}) scanning. Therefore there are out-of-focus maps labeled with ``\textit{L}'' and ``\textit{B}'', which correspond to single sets of three maps, and ``\textit{LB}'' that have been reduced using basket-weaving of two sets. The \textit{LB} sets build an average over a wider range of elevations, but usually have better signal-to-noise ratio. A linear baseline was subtracted from the single pass maps (\textit{L, B}), and a calibration factor of \qty{8}{\K} was applied to convert to antenna temperature (known value of the noise diode), and finally a sensitivity of $\Gamma = \qty{0.8}{\K\per\jansky}$ was applied, see \cref{eq:sensitivity}. After this process a total of 19 OOF holography set maps were obtained.

  The observation was performed by using the active surface turned {on}. The look-up table contains corrections up to \qty{1.5}{\mm} and without using it the maps would have been considerably degraded.
  There were \num{5} actuators (out of \num{96}) broken at the time of observation. Fortunately most of the broken actuators were located in the external ring (at radius \qty{3250}{\mm}; \cref{fig:actuators_sub-reflector}), where it is known that the polynomial fit does not reach good values (see \cref{fig:phase_sim_com}).

  \subsection{December 2020 observation campaign} \label{sec:ObsDec2020}

\begin{figure*}[h]
    \centering
    \includegraphics{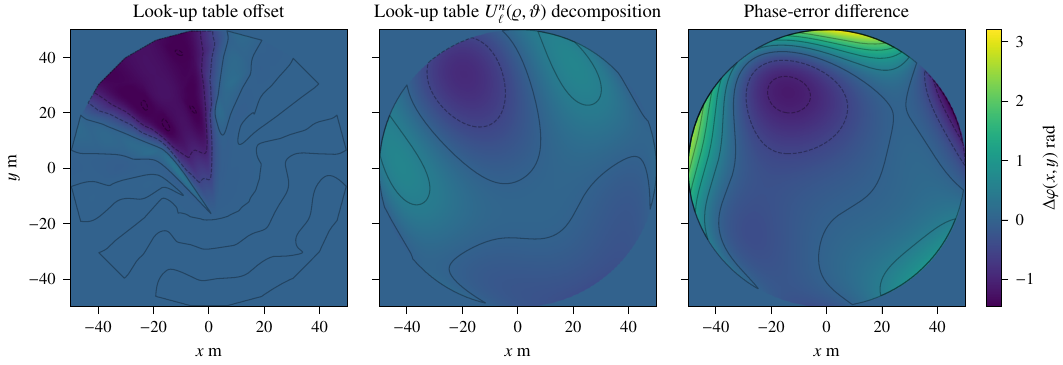}
    \caption{Closure and validation in orientation in FEM active surface control system. Observations were performed before and after applying an offset to the active surface, left-hand side. The right-hand side shows the difference between them. The middle figure shows the Zernike circle polynomials decomposition of the modified active surface. The contour lines are between -2 to 2 half-radian intervals.}
    \label{fig:phase_convention_7mm}
  \end{figure*}

  Further observations were done on, December \nth{16} and \nth{17} 2020, with the \qty{7}{\mm} receiver. These observations aimed to validate once again the true orientation of the look-up table and the phase-error maps computed using \texttt{pyoof} (similar to \cref{sec:ObsAug2017}).
  This time we observed with the active surface turned {on} while adding an extra offset to the FEM look-up table (left-hand side \cref{fig:phase_convention_7mm}). Additional observations were also performed in the same manner as in \cref{sec:ObsDec2019}, i.e., with the active surface {on} and standard look-up table.
  The idea is similar to what was done in \citet{2007A&A...465..685N}, who had a known offset in the FEM system and then retrieved it with OOF holography observations.

  The procedure is as follows: several observations with the same modified active surface (added offset) and similar elevation are averaged together (i.e., only averaging the resulted phase-error maps). Then, observations with roughly the same elevations as before, but with the standard active surface (no offset), are averaged together. Finally, the two averaged phase-error maps are subtracted and compared with the manually added offset from the look-up table. The added offset is applied to roughly a quarter of the \num{96} actuators. \Cref{fig:phase_convention_7mm}, from left to right, shows the applied offset in the grid of the look-up table, the same table decomposed in terms of Zernike circle polynomials (up to $n=5$), and finally the difference between the two: modified minus standard (active surface) phase-error maps. This observation is the final proof needed for the true orientation between the phase-error maps and the look-up table.

  \subsection{February and March 2021 observation campaign} \label{sec:ObsFebMar2021}

  Observations from early 2021 (March \nth{29} for 3C273 and three days in February \nth{11}, \nth{17}, and \nth{18} for 3C84) were used to cover a broader range in elevation of phase-error maps, which is particularly important in order to model repeatable elastic properties of the telescope mechanical structure (see \cref{sec:gravitational_model}). Observations were performed with the \qty{7}{\mm} receiver and the active surface turned on. There were still \num{4} out of \num{96} actuators broken but that should not affect the observations significantly. The procedure followed was that of the standard OOF holography observations, i.e., three beam maps to obtain one phase-error map.

  \subsection{July 2021 aperture efficiency observation}
  \label{sec:ObsJuly2021}
  The final set of observations was performed on July \nth{28}, 2021 in order to test the new look-up table (see \cref{sec:surface_improvements}) in contrast to the default FEM look-up table. In this case we did not perform the standard OOF holography observations, but instead observed three calibrators 3C273, 3C286, and 3C345; which are suitably bright at \qty{34.75}{\giga\hertz} with the \qty{7}{\mm} receiver. With these three sources a broad range in elevation could be covered and the observations were performed using cross-scans in azimuth and elevation. Cross-scans in azimuth and elevation are performed by slewing the antenna in the corresponding axis over the source position and recording the total power output of the receiver. The resulting data shows the beam profile and is well suited to measure the flux density of point like sources. An improved setting of the active surface should result in a higher efficiency and therefore an increase of the amplitudes of the cross-scans should be visible. 

  The cross-scans were reduced following the standard procedure for continuum flux density measurements \citep{2003A&A...401..161K}. The gain and amplitude calibration for all data is based on the scans that were measured with the original FEM look-up table. If the efficiency of a (new) OOF holography look-up table is better, the following scans should appear brighter and provide an easy and fast way to directly compare the different look-up tables (see \cref{sec:performance_of_the_oof_holography_look-up_table} for results).

  From all observations only higher signal-to-noise ratio $\geq\num{200}$ were used for analysis in \cref{sec:results}. A summary of all observation campaigns and their purposes is presented in \cref{tab:summary_obs}.

  \begin{table*}
    \centering
    \caption{Summary all observational campaigns. Column label with Maps corresponds to the number of phase-error maps computed from a single set of OOF holography observations (\num{3} beam maps) and each of them take about \qty{\sim45}{\minute}. The total number of maps gathers all types of scanning (azimuth, elevation, or combined) and also tests required to improve the \texttt{pyoof} software or find the true orientation between look-up table and phase-error maps. Only beam maps with a signal-to-noise ratio $\geq\num{200}$ were used for analysis in \cref{sec:results}. The Receiver column shows the receiver name convention at Effelsberg.}
    \label{tab:summary_obs}
    \begin{tabular}{llllll}
      \hline
      \hline
      Campaign date & Purpose & Maps & Receiver & Frequency & Elevation range \\ \hline
      August 2017 & Development \texttt{pyoof} and first orientation tests & - & \qty{9}{mm} & \qty{32.00}{\giga\hertz} & \qtyrange{69}{76}{\deg} \\
      December 2019 & Gather phase-error maps and test different receiver & 19 & \qty{7}{\mm} & \qty{34.75}{\giga\hertz} & \qtyrange{34}{78}{\deg} \\
      December 2020 & Gather phase-error maps and solving orientation & 10 & \qty{7}{\mm} & \qty{34.75}{\giga\hertz} & \qtyrange{61}{79}{\deg} \\
      February 2021 & Gather phase-error maps multiple elevations & 14 & \qty{7}{\mm} & \qty{34.75}{\giga\hertz} & \qtyrange{28}{69}{\deg} \\
      March 2021 & Gather phase-error maps multiple elevations & 10 & \qty{7}{\mm} & \qty{34.75}{\giga\hertz} & \qtyrange{27}{40}{\deg}\\
      July 2021 & Evaluate corrections to look-up table and closure & - &\qty{7}{\mm} & \qty{34.75}{\giga\hertz} & \qtyrange{17}{79}{\deg} \\
      \hline
    \end{tabular}
  \end{table*}

  \section{Results}
  \label{sec:results}
  
  In this section we present the analysis of OOF holography observations performed during six campaigns: the first in mid 2017 to test and analyze the conventions for the active surface and the aperture distribution; the second at the end of 2019, to test implementation of new \qty{7}{\mm} receiver; a third in late 2020 to validate and confirm the true convention in the active surface system and the \texttt{pyoof} phase-error maps; a fourth and fifth campaign to gather OOF holography phase-error maps at multiple elevations; and finally observations and tests of a new lookup table.

  \subsection{Random-surface-error efficiency}
  \label{sec:random-surface-error_efficiency}
\begin{figure}[t]
    \centering
    \includegraphics{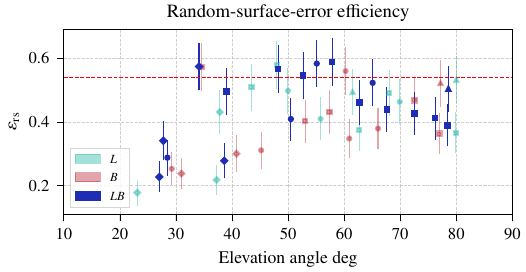}
    \caption{Random-surface-error efficiency ($\varepsilon_\text{rs}$) as  a function of the elevation ($\alpha$) for the primary dish. 
    $\varepsilon_\text{rs}$ corresponds only to the large-scale deviations found by the OOF holography method.
    Each phase-error map was calculated with $n=5$ and then compute the efficiency using \cref{eq:e_rs,eq:displacement}.
    The dashed red line represents the traditionally defined value for a good efficiency, $\delta_\text{rms}\leq{\lambda}/{16}$ $(\varepsilon_\text{rs}=0.54)$. The labels \textit{L} and \textit{B} are azimuth and elevation single scans, and \textit{LB} are the combination in the basket-weaving mode. Squares are December 2019,  triangles December 2020, circles February 2021, and diamonds March 2021 observations.}
    \label{fig:epsilon_rse}
  \end{figure}

  Each phase-error map, $\varphi^{(\alpha)}(x, y)$, can be used to compute the random-surface-error efficiency ($\varepsilon_\text{rs}$; same as in eq.~\ref{eq:e_rs}) at a mean elevation, $\alpha$. \Cref{fig:epsilon_rse} shows the random-surface-error efficiency for the complete set of observation campaigns, differentiating the azimuth, elevation and basket-weaving methods ($L$, $B$, and $LB$ respectively). Maps that undergo a basket-weaving will have, in general, a higher signal-to-noise ratio, but a less constrained elevation angle.
  Error bars were estimated from the known mean surface error of the Effelsberg reflectors. The dashed red line shows the traditionally good value for surface deviations, $\delta_\text{rms} \leq \frac{\lambda}{16}$ \citep[see][Chapter~7]{stutzman1998antenna}. \Cref{fig:epsilon_rse} shows a clear deviation with respect to elevation, and a lower level from $\varepsilon_\text{rs}=0.54$. From that we can state that there is room for improvements in the overall optical system and that by correcting the look-up table we can improve $\varepsilon_\text{rs}$. The need for a better elevation and reflector corrections is also visible in the most commonly used aperture efficiency \cref{eq:aperture_efficiency}, usually computed by observations where the dependence to other efficiencies cannot be easily decoupled.
  
  Notice that the $\varepsilon_\text{rs}$ calculation is highly dependent in the \text{signal-to-noise ratio} of the source (i.e., flux density at these frequencies; \cref{tab:sources}). Lower elevation observations (March 2021) were performed on 3C273, which had a significantly lower flux density than 3C84 at the time of the observations. Especially the March 2021 observations (diamond symbols in \cref{fig:epsilon_rse}) were close to the $\text{signal-to-noise ratio}\approx\num{200}$ threshold.

  \subsection{Gravitational model}
  \label{sec:gravitational_model}

  As discussed in \cite{2007A&A...465..685N}, one can measure the elasticity properties due to the gravitational effects depending on elevation through the Zernike circle polynomial coefficients, $K_{n\,\ell}(\alpha)$. 
  Unfortunately the only way to perform OOF holography observations at Effelsberg requires that the current active surface (FEM model look-up table) had to be turned on during the entire observations (otherwise degradation in the beam maps would have reduced $\text{signal-to-noise ratio} \leq200$), hence when applying corrections to the look-up table a mix between the current FEM model and phase-error maps will be used. This is done by simply applying the difference between the FEM look-up table (transformed to phase-error, eq.~\ref{eq:displacement}) minus the observed (OOF holography) phase-error:
  \begin{equation}
    \varphi^{(\alpha)}_\text{C} = \varphi^{(\alpha)}_\text{FEM} - \varphi^{(\alpha)}_{\text{OOF}},
    \label{eq:fem-oof}
  \end{equation}
  where $\varphi_\text{C}$ is the final correction done to the look-up table, $\varphi_\text{FEM}$ the default FEM phase values (standard look-up table), and $\varphi_{\text{OOF}}$ is the obtained solution from OOF holography while observing with the standard FEM look-up table (active surface turned on).

  Now taking the 18 coefficients sets for each OOF holography map observed from December 2019, December 2020, February 2021, and March 2021; we can estimate a gravitational model \citep{2007A&A...465..685N} for the Effelsberg telescope. By considering only those deformations by gravity, which are large-scale, well constrained, repeatable, and in a long enough time scale to correct, we can decompose the gravity vector as:
  \begin{equation}
    K_{n\,\ell}(\alpha) = g_{n\,\ell}^{(0)} \sin\alpha + g_{n\,\ell}^{(1)}\cos\alpha + g_{n\,\ell}^{(2)}.
    \label{eq:gravity}
  \end{equation}
  The $g_{n\,\ell}^{(i)}$ ($i=0, 1, 2$) coefficients are unique for a single Zernike circle polynomial coefficient, $K_{n\,\ell}$. In principle, $K_{n\,\ell}=K_{n\,\ell}\del{\alpha}$ is a function of the elevation, so a linear fit can be applied to find the missing $g_{n\,\ell}^{(i)}$ set. The computations of $K_{n\,\ell}$ are performed only on the maps $\varphi^{(\alpha)}_\text{OOF}$, \cref{eq:fem-oof}, which will have some gravitational dependency not removed by FEM.

\begin{figure*}[t]
    \centering
    \includegraphics{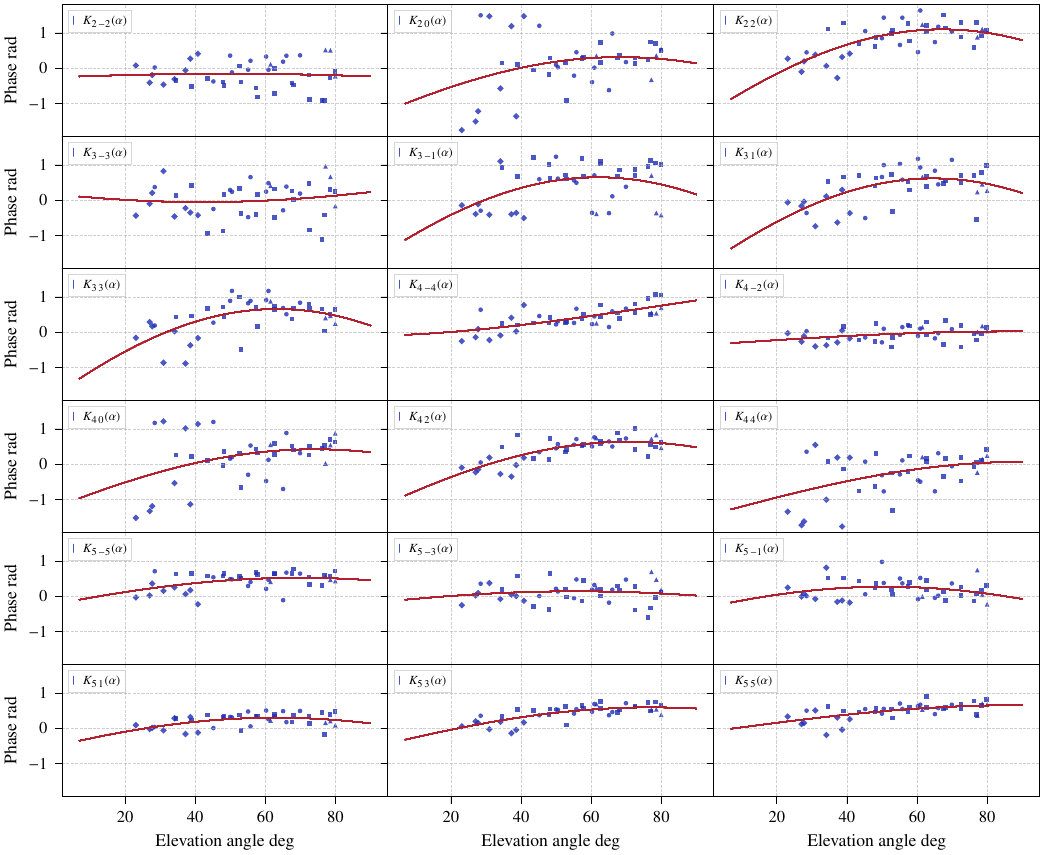}
    \caption{Gravitational model in terms of Zernike circle polynomial coefficients. The squares, triangles, circles, and diamond represent OOF holography observations from December 2019, December 2020,  February 2021, and March 2021; respectively. The red line is the best fit made to the data using \cref{eq:gravity} with three $g_{n\,\ell}^{(i)}$ coefficients. A flat model indicates that no corrections are needed to the $K_{n\,\ell}$ coefficient.}
    \label{fig:gnl_fit}
  \end{figure*}

  \Cref{fig:gnl_fit} shows the result of the gravitational model applied to the observations from December 2019, December 2020, February 2021, and March 2021. The plot is divided into the \num{19} $K_{n\,\ell}$ coefficients relevant to the phase-error. There is a clear trend shown by the red curve while increasing/decreasing the elevation, meaning there is still room for an improvement in the current look-up table (same as in \cref{fig:epsilon_rse}). Coefficient $K_{n\,\ell}$ uncertainties are measured from the covariance output in the least-squares minimization, where most measurements show an error percentage ${\delta K_{n\,\ell}}/{K_{n\,\ell}}\approx\num{3.3e-2}$.

\begin{figure*}
    \centering
    \includegraphics{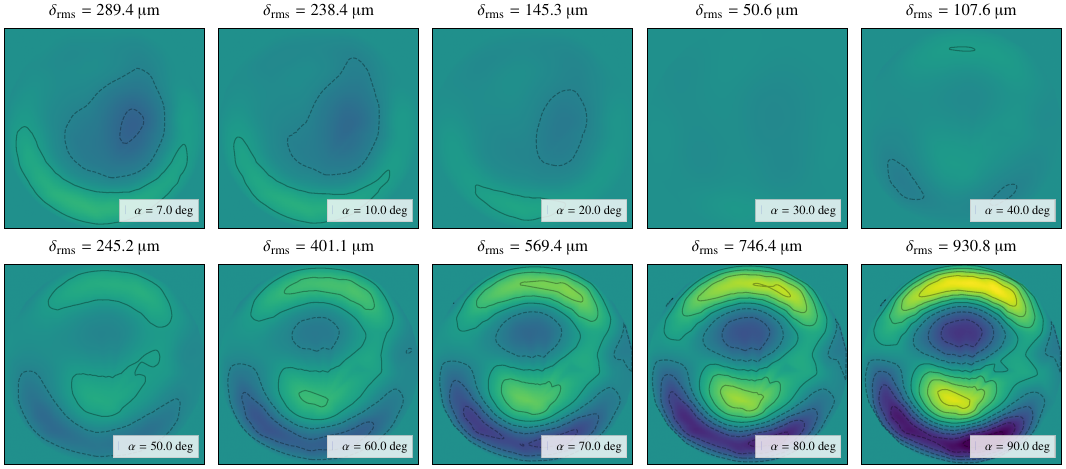}
    \caption{Computation of the new look-up table with OOF holography and FEM corrections. Each panel shows the expected amplitude ${\varphi_\bot}_\text{C}$ per elevation $\alpha$ with its rms value, $\delta_\text{rms}$. The table is based on the gravitational model (shown in \cref{fig:gnl_fit}) and default FEM look-up table (\cref{fig:fem_lookup}). The table is a result of multiple OOF holography observations at several elevations in order to create a reliable gravitational model (eq.~\ref{eq:gravity}). The contour lines are between \qtylist{-2000;2000}{\micro\m}, with \qty{400}{\micro\m} intervals. The values determined here correspond well to the observed main dish deformations by \citet{2015JAGeo...9....1H}.}
    \label{fig:new_lookup}
  \end{figure*}

  The new table computed from gravitational deformations follows \cref{eq:aperture_phase_distribution,eq:displacement,eq:fem-oof} in order to find the correspondent actuator amplitude, ${\varphi^{(\alpha)}_\bot}_\text{C}$. The full elevation set of the look-up table is shown in \cref{fig:new_lookup}. Each elevation corresponds to \num{96} actuators entries in the active surface system.
  Notice that most of the corrections are based on the original FEM look-up table (see \cref{ap:effelsberg_active_surface_control_system} and \cref{fig:fem_lookup}), nonetheless, there is still room for improvement.

  \subsection{Performance of the OOF holography look-up table}
  \label{sec:performance_of_the_oof_holography_look-up_table}

  To finally test the improved performance of the OOF holography look-up table (from the computed gravitational model in \cref{fig:gnl_fit}), the direct relative gain $G_\text{rel}$ (proportional to the aperture efficiency) was computed from three calibration sources (\cref{sec:ObsJuly2021}). The aperture efficiency is not so easy to compute, instead we use the telescope's sensitivity (eq.~\ref{eq:sensitivity}) and flux density of the source at a given elevation.

\begin{figure*}[t]
    \centering
    \includegraphics{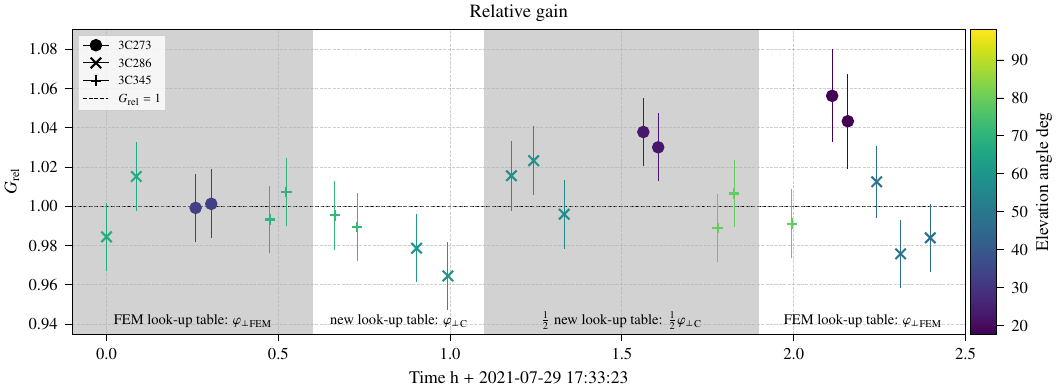}
    \caption{The relative gain for the selected sources at different elevations observed with the original FEM model actuator table, and with the modified tables. The error bars of the data points combine the individual error of the Gaussian fit to the cross-scans and the calibration uncertainty of the main calibrator 3C286. For the modified tables the full corrections calculated by the OOF holography (gravitational model) were applied and as a test also only half of the amount of the corrections were applied. The different scenarios are separated by the gray region: FEM look-up table (default) ${\varphi_\bot}_\text{FEM}$, new look-up table ${\varphi_\bot}_\text{C}$, and half new look-up table ${\varphi_\bot}_\text{C}/2$. The dashed horizontal line shows a relative gain of $G_\text{rel}=\num{1}$, corresponding to no difference to the FEM model.}
    \label{fig:oofh_table_nflux}
  \end{figure*}

  Results can be seen in \cref{fig:oofh_table_nflux} where different scenarios were computed, i.e., using the standard FEM look-up table ${\varphi_\bot}_\text{FEM}$, the OOF holography look-up table ${\varphi_\bot}_\text{C}$ (eq.~\ref{eq:fem-oof}), and half OOF holography look-up table ${\varphi_\bot}_\text{C}/2$. The ${1}/{2}$ factor was introduced due to the known high amplitude values resulting from the gravitational model, especially at the edges of the phase-error (due to the less well constrained solution), which might cause an overestimation of the phase errors.
  
  At first glance the measurements with the OOF holography look-up table indeed show somewhat lower amplitudes than the original FEM model (first two regions from left in \cref{fig:oofh_table_nflux}). The measurements in the third region ($\frac{1}{2}{\varphi_\bot}_\text{C}$), however, are on average higher than the original FEM model data. This suggests that the corrections calculated from the OOF holography are correct, but might overestimate the phase-error amplitudes in some parts. It is known that the search for an optimized surface accuracy using OOF holography is an iterative process \citep{2007A&A...465..685N}. The last region in the figure shows again measurements with the original FEM model for comparison. The points at low elevation correspond to 3C273, which was then below $\qty{17}{\deg}$ elevation where measurements become more uncertain because of the higher air mass and the corresponding atmospheric corrections. In summary the results show that the theoretical FEM model for the Effelsberg antenna is already quite good and robust. The slightly higher aperture efficiency, $G_\text{rel}=G_\text{rel}\del{\varepsilon_\text{rs}}$, when considering the measurements at half-amplitude of the proposed corrections show that there might be some room for a few percentage improvement and we will continue to investigate this with more measurements.

\begin{figure}
    \includegraphics{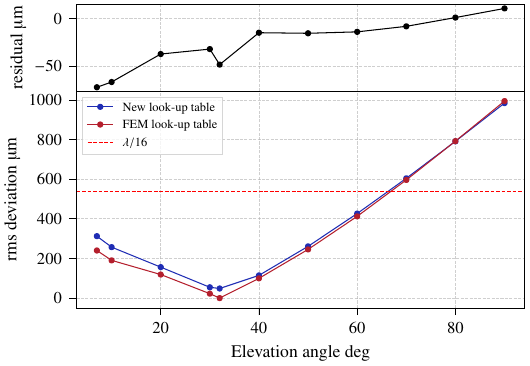}
    \caption{Root-mean-square (rms) deviation of the FEM ${\varphi_\bot}_\text{FEM}$ and new ${\varphi_\bot}_\text{C}$ look-up tables. The rms value for each case only contains the 96 actuators positions and not the entire computed phase-error map. Upper plot shows the difference from both methods. Most of the corrected deviation happens at lower elevations, where the random-surface-error efficiency dominates (see \cref{fig:epsilon_rse}). The horizontal line corresponds to the traditionally defined value for a good efficiency, $\delta_\text{rms}\leq{\lambda}/{16}$ $(\varepsilon_\text{rs}=0.54)$, and it has been drawn only as an indicator.}
    \label{fig:rms_elevation}
  \end{figure}

  To understand these empirical results in \cref{fig:oofh_table_nflux}, we can compare the FEM and the new look-up table rms in the given elevation steps. \Cref{fig:rms_elevation} shows the rms deviation of the FEM ${\varphi_\bot}_\text{FEM}$ and new ${\varphi_\bot}_\text{C}$ look-up tables. Their difference (upper panel \cref{fig:rms_elevation}) is very similar above elevation \qty{60}{\degree} and below \qty{40}{\degree} differences become evident. As seen in \cref{fig:oofh_table_nflux} lower elevations present the most significant improvement in relative gain, $G_\text{rel}$. \Cref{fig:rms_elevation} is also in agreement with the random surface error efficiency where most of the available room for improvement is at lower elevations.

  \section{Discussion} 
  \label{sec:discussion}

  \subsection{Parameters and polynomial fit}
  \label{sec:params_and_polynomial_fit}

  The OOF holography method has a set of free parameters that need to be addressed; among these, the most important are: aperture distribution model (number of coefficients to fit and orthonormality), axial offset magnitude added to the telescope ($d_z$), and number of beam maps.

  \subsubsection{Aperture distribution model}
  \label{sec:aperture_distribution_model}
  The model first introduced by \citet{2007A&A...465..679N} has been modified to improve the performance at the Effelsberg telescope (e.g.: telescope geometry, OPD, and coefficient algorithm/fit).

  From simulations in \cref{sec:simulations}, it was seen that the \texttt{pyoof} package performs a relatively good fit for the power pattern. What's more, several combinations of a limiting OOF holography signal-to-noise ratio were performed and the \texttt{pyoof} package was able to trace back a good approximation for the randomly generated $K_{n\,\ell}$ coefficients.

  The nonlinear least-squares minimization returns a series of useful information: power pattern residual, and variance-covariance and correlation matrices. All of them are directly computed from the determination of the solution vector $\boldsymbol\theta$. It is important to study the optimization result in order to not fall into an over-fitting situation. From the first observation campaigns and later simulations (\cref{sec:simulations}), it is clear that there is a high correlation between $A_{E_\text{a}}\leftrightarrow K_{0\,0}$ and $(x_0, y_0)\leftrightarrow(K_{1\,1}, K_{1\,-1})$ (tilt aberration) coefficients. Fortunately by normalizing the power pattern and setting to one and zero such coefficients, the degeneracies vanish. The next most degenerate coefficients are $c_\text{dB}$ and $q$, but after several observations they tend to repeat, hence both of them were fixed, to $c_\text{dB}=\qty{-14.5}{\decibel}$ and $q=1.4$.

  The Zernike circle polynomial coefficients are known to be orthonormal under the unitary circle, but at the time of the calculation, the aperture distribution, the blockage distribution, and the illumination function will affect the behavior of this orthonormality. This becomes more evident in the correlation matrix while increasing the polynomial order $n\geq6$.
  The solution is simply to restrict the order until $n=6$. Besides this, it is well known that up to $n=5$ the polynomials are well represented by classical aberrations\footnote{A more detailed description of the Zernike circle polynomials coefficients can be found in the \texttt{pyoof} package documentation \url{https://pyoof.readthedocs.io/en/latest/zernike/index.html}.} (defocus, astigmatism, coma, and spherical). Such aberrations also contain only large-scale aberration over the unitary circle, where the small and less well constrained aberrations become negligible. Even more, a better resolution in the phase-error maps will not improve the results at the Effelsberg telescope due to the limitation of the active surface itself. The minimum spacing between actuators in the sub-reflector is of the order of \qty{63}{\mm} which represents up to \qty{1}{\m} aberration in the primary dish (and it is only possible in the inner section of the dish). A more general description of the active surface control system is in \cref{ap:effelsberg_active_surface_control_system}.

  \subsubsection{Axial offset $d_z$}
  The axial offset is the key piece to retrieve the phase-error. This added term, gives the defocused shape in the beam maps (see \cref{fig:beam_sim_com} first and third column). 
  The signal-to-noise ratio of the defocused beam will depend on the $d_z$ magnitude, i.e., the higher the magnitude, the larger the shape of the defocused pattern, which implies a lower overall signal in the beam. Besides this effect, a larger mapping area $>\qty{340}{\arcsec}\times\qty{340}{\arcsec}$ may be necessary, i.e., extending the observing time ($\gtrsim\qty{45}{\minute}$) and add more uncertainty in the estimated mean elevation.
  On the other hand, a smaller $d_z$ magnitude will generate a smaller defocused beam that will not present evident structure to differentiate it from a focused beam, which implies less constraints in the fitting algorithm (larger correlation in the fitted parameters $\boldsymbol\theta$; eq.~\ref{eq:coeff_vector}). 
  Depending on the receiver, center frequency, and collecting area, $d_z$ needs to be adjusted, see \cref{tab:obs_config}, and it will be dependent on the telescope-receiver configuration. At Effelsberg this has been an empirical process to optimize signal-to-noise ratio and results from the fitting algorithm.

  \subsection{Phase-error and FEM look-up table convention}
  \label{sec:phase-error_and_fem_look-up_table_convention}

  In order to study surface deviations and correct them, we need to understand conventions in the phase-error maps (\texttt{pyoof}) and FEM look-up table. As seen in \cref{fig:phase_convention_9mm,fig:phase_convention_7mm}, using a \qty{9}{\milli\metre} and \qty{7}{\mm} receiver, respectively, we are able to find the true orientation. Limitations in finding the true orientation are: poorly constrained edges (mainly due to the illumination function) and the reduced number of actuators (\num{96} at Effelsberg, \cref{fig:actuators_sub-reflector}). Nevertheless, we can find clear evidence while performing observations with the active surface turned off and retrieve the expected FEM behavior (\cref{fig:phase_convention_9mm}) and by observing with a known offset in the active surface and retrieving it back again (\cref{fig:phase_convention_7mm}).

  \subsection{Validity of the OOF holography method}
  \label{sec:validity_of_the_oof_holography_method}

  The exposed analyses of the method applied to the Effelsberg telescope can be summarized as follows:
  
  \begin{itemize}
    \item We have shown that the OOF holography method applied to the Effelsberg 
    telescope works as expected. 

    \item Simulations showed that the numerical solver works as expected and is capable of retrieving deliberately added offsets (\cref{sec:simulations}). The validity of the used conventions concerning the telescope optics could be proved as well (\cref{sec:ObsAug2017,sec:ObsDec2020}).
    
    \item From a set of systematic observations a quantitative measurement of the surface aberrations can be obtained, indicating an elevation dependency (hence, potential for improvement; \crefrange{sec:ObsDec2019}{sec:ObsFebMar2021}).
    
    \item A gravitational model has been obtained from the observing campaigns and a new look-up table for the FEM system was computed (\cref{sec:gravitational_model}).
    
    \item The look-up table was updated and tested resulting in a modest improvement 
    (expressed in terms of a relative gain) mostly at elevations $\alpha < \qty{30}{\degree}$. That proved that the original FEM model gave already a proper description of the main reflector (\cref{sec:gravitational_model,fig:oofh_table_nflux,fig:rms_elevation}).

    \item Limitations of this analysis are discussed in \cref{sec:improved_look-up_table_and_limitations}--- addressing these will likely lead to further improvements.
  \end{itemize}

  \subsection{Improved look-up table and limitations}
  \label{sec:improved_look-up_table_and_limitations}

  A gravitational model as suggested in \citet{2007A&A...465..685N} was developed for the Effelsberg telescope in \cref{sec:gravitational_model} and shows results of a clear elevation dependency for the Zernike circle polynomials coefficients. \Cref{fig:gnl_fit} shows the trend of the best fit where in most cases the scattered data correctly follows the line, with the exception of $K_{2\,0}$ (defocus), and $K_{4\,0}$ (primary spherical). These large deviations can be blamed on the active surface itself, where the distribution of actuators lacks in the center of the aperture plane (\cref{fig:actuators_sub-reflector}, first ring \qty{1210}{\mm}). 

  Clear trends as a function of elevation in \cref{fig:epsilon_rse,fig:gnl_fit} show that there exists room for improvements to the current table and that the OOF holography method could be implemented at Effelsberg.

  A new look-up table $\del{{\varphi_\bot}_\text{C}}$ is then recalculated using OOF holography and FEM corrections combined, and tested with relative gain in \cref{sec:performance_of_the_oof_holography_look-up_table}. Results with the proposed solution do not show a clear improvement in \cref{fig:oofh_table_nflux}. 
  Assuming $\varepsilon_\text{ph}\sim\varepsilon_\text{rs}$ we then expect a one-to-one increase in $G_\text{rel}$ after correcting by surface deviations (\cref{fig:epsilon_rse}).

  The lack of stronger corrections in \cref{fig:new_lookup} compared to \cref{fig:fem_lookup} at high elevations (\cref{fig:rms_elevation}) is likely due to a combination of several factors. 
  Limiting signal-to-noise ratio (lower flux density) for some of the observations, particularly at low elevations (\cref{sec:random-surface-error_efficiency}).
  The present correlation between some illumination and Zernike circle polynomial coefficients (not fully orthonormal solution).
  Since an astronomical receiver is used, the tapering of the edges (to avoid spill over) does not allow to well constrain the values at the phase-error edges. 
  In addition, the OOF holography observations were scattered over many months/years and therefore have been made at different conditions in terms of thermal gradients, wind loads, and sometimes a small number of actuators were broken.
  An important constraint is that the improvements applied to the active surface of the sub-reflector have a limited resolution and correspond to rather large scale deformations of the main dish (spread over \qty{\sim10}{\m}).
  Nevertheless, our results show that the current FEM model is already a very good representation of the main dish's properties.

  \subsection{General considerations and applicability to other radio telescopes}
  \label{sec:general_considerations_and_applicability_to_radio_telescopes}

  To implement the OOF holography method in another facility, and considering at least 3 beam maps (one in- and two out-of-focus), the following considerations must be taken into account.

  \begin{itemize}
    \item \textbf{Telescope geometry}: this will dictate the OPD $\delta\del{x,y;d_z}$ which will depend on the type of telescope optics (can also be a numerical model; \citealt{2018ITAP...66.2044D}), and blockage function $B\del{x,y}$ to use (including telescope struts and sub-reflector is crucial).
    
    \item \textbf{Illumination function} $E_a\del{x,y}$: taper model with its estimated parameters to decouple it from the $K_{n\,\ell}$ set. Ideally these parameters should not vary, keeping them fixed will reduce parameter degeneracies.
    
    \item \textbf{Observing frequency} $\nu$: In principle it is true that the higher the frequency, the more accurate the phase-error distribution will be. However, the efficiency of antennas reduces at higher frequencies and typical astronomical sources get weaker. In addition, the negative effects of the atmosphere become larger and therefore accurate observations at higher frequencies are more difficult to obtain.
    Multi-frequency receivers may be particularly useful since their information can be added to improve the nonlinear least-squares optimization (not available at the time at Effelsberg but implemented in \citealt{2007A&A...465..685N}). Higher frequencies may not reveal higher aberration orders since they are strictly dependent in the number of Zernike circle polynomials used (and convergence of the method).
    
    \item \textbf{Axial offset} $d_z$: We observed that our instrument and software had a good response while having a $d_z = \pm 2.2\lambda$ value (although GBT used a $d_z=\pm5\lambda$ to fit two defocused beams; \citealt{2007A&A...465..685N}). This was empirically tested over a trial and error procedure, and it was a compromise among: defocused beam size $d_z$, scanning time, and signal-to-noise ratio.
    
    \item \textbf{Size of the map}: The size is coupled with the defocused beam pattern. The map has to cover at least the full beam pattern plus some \qty{10}{\percent} more area on both sides to allow a proper removal of baseline instabilities due to weather effects. Here a compromise has to be found. Larger defocus ($d_z$) will cause a larger beam map and a larger beam map takes longer to perform. Since the elevation of the antenna changes with time and therewith the gravitational deformations change it is sensible to keep the observing time short and still have enough defocus to properly fit the beam pattern.
    
    \item \textbf{Number of scanning points} (sampling): The number of scanning points used will be given by the size of the map and the observing frequency. We mapped the point source with steps \num{\sim.3} the size of the telescope beamwidth (to remove pixel effects; e.g., Nyquist sampling along the diagonal), e.g., with this constraint \cref{fig:beam_obs_com}: observing at \qty{34.75}{\giga\hertz} and a map size of \qty{\sim0.1}{\deg} we get \num{\sim39}; $\num{39}\times\num{39} = \num{1521}$ for a square grid.
  \end{itemize}
  
  \subsection{Future considerations}
  As discussed in \cref{sec:aperture_distribution_model}, the Zernike circle polynomials are not orthonormal under the telescope's aperture, and for this to be, we would need to consider a new set of orthonormal polynomials with the illumination function and blockage distribution response included. Such a case has already been developed in optical astronomy with the adaptive optics (AO) methods, where in contrast to radio wavelengths the focus is in correcting for rapid variations in the Earth's atmosphere over \qty{\sim1}{\milli\s} timescales (faster than the telescope's slew). Some AO systems take into consideration non-circular apertures with complex blockage distributions, but nevertheless a set of orthonormal polynomials can be computed numerically. 
  The first one of these corrections developed by \citet{1984JOSAA...1..685M} (analytic solution) shows an orthonormal set capable of dealing with the sub-reflector blockage (most radio telescopes), but a more complete and generalized approach has been developed by \citet{1994ApOpt..33.1832S} and \citet{2004OptL...29.2840U} (numerical solutions). In general, this approach will be the best possible, but prior to the orthonormal numerical set construction, the illumination function and blockage distributions must be known; on the contrary such a numerical set could fail to retrieve a realistic value for aberrations. Restrictions in degrees of freedom will imply a change in the orthonormal set, making the problem highly specific for a particular telescope and resulting in a longer time needed to find the correct orthonormal set to be used.

  The mechanical structure non-repetitive effects such as the temperature and wind load were not discussed in the presented work, due to the long timescale of the performed OOF holography measurements. In order to study such fluctuations measurements need to be at least of the same time order of the acting load. A fast observing method could also be considered for future observations.
  Other considerations in the stability of a single set of Zernike circle polynomial coefficients (decomposed into $g_{n\,\ell}^{(i)}$) for the Effelsberg telescope have not been included in the presented work. Measurements at different hours of the day may impact the mechanical structure differently due to thermal gradients, making an OOF holography table valid only for a specific period of time.

  \section{Conclusions} 
  \label{sec:conclusions}

  A new software, \texttt{pyoof}, for OOF holography was developed and specialized for the Effelsberg telescope, as well as its expansion and documentation for use at other facilities. The software is able to retrieve the aperture phase distribution given three standard OOF holography beam maps in order to break the degeneracy between the power pattern and the aperture distribution. An early stage for an active surface implementation, specifically for Effelsberg, has now been implemented (and in principle adaptable to other facilities).

  Systematic OOF holography observations were performed at the Effelsberg telescope and analyzed over four years. Among the most prominent results are the clear correspondence of classical aberrations and observed phase-error maps (\cref{sec:ObsAug2017}), the correct optical orientation of the phase-error and active surface (\cref{sec:ObsDec2020}), the computation of a gravitational model ($g_{n\,\ell}^{(i)}$ coefficients, \cref{sec:gravitational_model}), and modest performance of the aperture efficiency with a new look-up table ${\varphi_\bot}_\text{C}$ (\cref{sec:performance_of_the_oof_holography_look-up_table}).
  The latter does not show a clear improvement in the overall telescope's aperture efficiency, and further corrections are needed in order to solve the ${1}/{2}$ factor observed in the corrected look-up table ${\varphi_\bot}_\text{C}$. Nevertheless, results are in agreement with the FEM model and further validate its robustness. 

  OOF holography measurements can be implemented at any telescope following \texttt{pyoof} and the listed indications throughout this article. Further considerations can also be included programmatically and in practice to boost the current performance of the method and adapt it to other instruments.

  Lastly, we emphasize the importance of developing holographic methods in radio astronomy, particularly for higher radio frequencies (e.g., \citealt{2022SPIE12190E..07R}) where surface accuracy is critical, and access to holographic information is limited in most cases at a fixed elevation. Without a constant monitoring of telescope optics, what is currently being done at optical facilities, degradation of radio signals will be inevitable.

  \begin{acknowledgements}
    We wish to thank the staff at the Effelsberg 100-m telescope for their support during the observations. The Effelsberg 100-m telescope is operated by the Max-Planck-Institut f\"ur Radioastronomie (MPIfR).

    We thank the developers of \texttt{astropy} \citep{2013A&A...558A..33A,2018AJ....156..123A}, \texttt{numpy} \citep{2020arXiv200610256H}, \texttt{scipy} \citep{2020NatMe..17..261V}, \texttt{matplotlib} \citep{2007CSE.....9...90H}, and the first OOF holography software by \cite{2007A&A...465..679N}.

    Finally, T.~Cassanelli wishes to thank Prof.~K.~Menten, for his support and invaluable help during the past years.
  \end{acknowledgements}

  \bibliographystyle{aa.bst}          
  \bibliography{oofh_effelsberg.bib}  

  \begin{appendix}
    \section{Zernike circle polynomials}
    \label{ap:zernike_circle_polynomials}

\begin{figure*}
      \centering
      \includegraphics{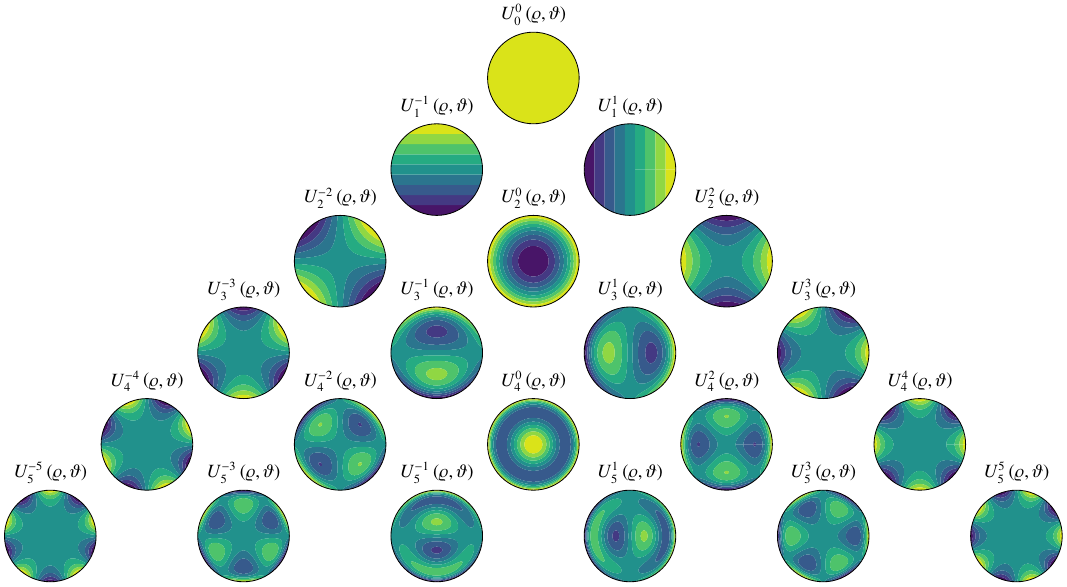}
      \caption{Zernike circle polynomials. This is the polynomial and naming convention used in all the presented analysis. Largest amplitude values are yellow and lower are purple. Amplitudes oscillate between one and minus one. The Zernike circle polynomials $U^\ell_n(\varrho, \vartheta)$ are up to a radial order of $n=5$, i.e., 21 polynomials.} 
      \label{fig:zernike_convention}
    \end{figure*}

    The Zernike circle polynomials are a set of orthonormal polynomials under the unitary circle. Although there are many sets of polynomials that gather this property, only the Zernike circle polynomials preserve simple properties of invariance \citep[see][Chapter~9.2]{1965poet.book.....B}.
    Depending on the convention the polynomials can have different orientations and relative order \citep{2022JOpt...24l3001N}; in the presented analysis we have used the following order from top to bottom in \cref{tab:zernike_aberration}, a polynomial definition from \citeauthor{1965poet.book.....B} and sorted same as Malacara (with no prior normalization). The definition from \citeauthor{1965poet.book.....B} follows: $0\leq\varrho\leq 1$ and $\vartheta$ increases from the $+x$ anticlockwise.
    \Cref{fig:zernike_convention} plots these same polynomials in the right $x$- and $y$-axis orientation to form the phase-error maps in the \texttt{pyoof} software.
    
    Other notations use the $j$-index and omit the $\ell$ classic formalism from \citep{1965poet.book.....B}.
    \begin{equation}
      Z_j\del{\varrho, \vartheta} = Z^m_n\del{\varrho, \vartheta} = \begin{cases}
        R^{\envert{m}}_n\del{\varrho} \cos\del{m\vartheta}, & m\geq 0, \\
        R^{\envert{m}}_n\del{\varrho} \sin\del{\envert{m}\vartheta}, & m<0.
      \end{cases}
    \end{equation}
    Nevertheless, consistency in the ordering, normalization, and orientation of the polynomials must return same solutions.

    \begin{table}[t]
      \caption{Low order Zernike circle polynomials and their correspondent name in aberrations. Same order is used in the \texttt{pyoof} software to find and fit the $K_{n\,\ell}$ constants.}
      \label{tab:zernike_aberration}
      \begin{tabular}{rrlp{3.25cm}}
        \hline
        \hline
        $n$ & $\ell$ & {Polynomial} $U^\ell_n(\varrho,\vartheta)$ & {Name} \\ \hline
        \num{0} & \num{0} & \num{1} & Piston \\
        \num{1} & \num{-1} & $\varrho\sin\vartheta$ & $y$-tilt \\
        \num{1} & \num{1} & $\varrho\cos\vartheta$ & $x$-tilt \\
        \num{2} & \num{-2} & $\varrho^2\sin2\vartheta$ &  
        \qty{45}{\degree} primary astigmatism \\
        \num{2} & \num{0} & $(2\varrho^2-1)$ & Defocus \\
        \num{2} & \num{2} & $\varrho^2\cos2\vartheta$ & \qty{0}{\degree} primary astigmatism \\
        \num{3} & \num{-3} & $\varrho^3\sin3\vartheta$ & \\
        \num{3} & \num{-1} & $(3\varrho^3-2\varrho)\sin\vartheta$ & Primary $y$-coma \\
        \num{3} & \num{1} & $(3\varrho^3-2\varrho)\cos\vartheta$ & Primary $x$-coma \\
        \num{3} & \num{3} & $\varrho^3\cos3\vartheta$ & \\
        \num{4} & \num{-4} & $\varrho^4\cos4\vartheta $ & \\
        \num{4} & \num{-2} & $(4\varrho^4-3\varrho^2)\sin2\vartheta$ & \qty{45}{\degree} secondary astigmatism \\
        \num{4} & \num{0} & $(6\varrho^4-6\varrho^2+1)$ & Primary spherical \\
        \num{4} & \num{2} & $(4\varrho^4-3\varrho^2)\cos2\vartheta$ & \qty{0}{\degree} secondary astigmatism \\
        \num{4} & \num{4} & $\varrho^4\cos4\vartheta $ & \\
        \num{5} & \num{-5} & $\varrho^5\sin5\vartheta$ & \\
        \num{5} & \num{-3} & $(5\varrho^5-4\varrho^3)\sin3\vartheta$ & \\
        \num{5} & \num{-1} & $(10\varrho^5-12\varrho^3+3\varrho)\sin\varrho$ & Secondary $y$-coma \\
        \num{5} & \num{1} & $(10\varrho^5-12\varrho^3+3\varrho)\cos\varrho$ & Secondary $x$-coma \\
        \num{5} & \num{3} & $(5\varrho^5-4\varrho^3)\cos3\vartheta$ & \\
        \num{5} & \num{5} & $\varrho^5\cos5\vartheta$ & \\
        \hline
      \end{tabular}
    \end{table}

    \section{Effelsberg telescope geometry}
    \label{ap:effelsberg_telescope_geometry}
    The Effelsberg telescope is a 100-m dish located in Bad M\"{u}nstereifel, Germany. It was finished in 1972 and until 2000 it was the largest fully-steerable single-dish telescope available. Its most important construction principle is that of the ``homologous deformation'',
    meaning that the main dish is elastic and when being tilted always maintains a parabolic shape \citep{1973IEEEP..61.1288H}. \Cref{tab:effelberg_geometry} shows some basic geometrical dimensions of the telescope \citep{2014JGeod..88.1145A}.

    \begin{figure*}
      \centering
      \includegraphics{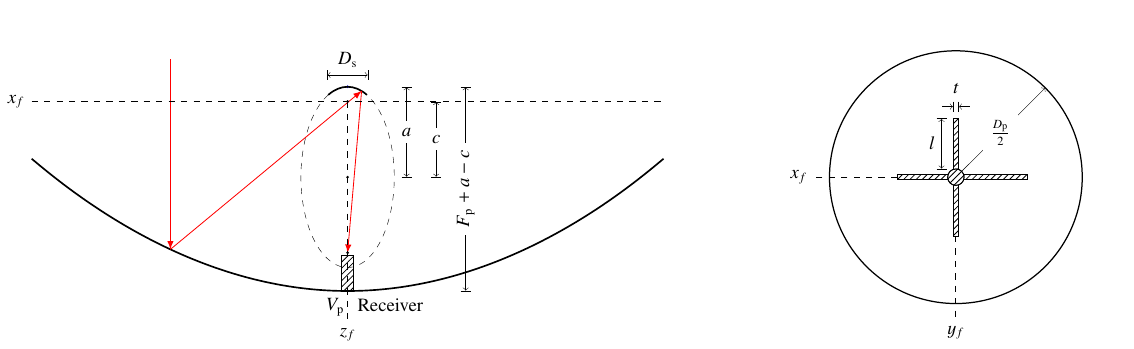}
      \caption{Effelsberg Gregorian telescope. Left: cross section $x_fz_f$-plane  primary reflector (parabola) and sub-reflector (ellipse). Right: aperture plane or $x_fy_f$-plane projection, the dashed area corresponds to the support struts and the sub-reflector (part of the blockage distribution, see \cref{fig:blockage}), see \cref{tab:effelberg_geometry}. The receiver or feed antenna is placed in the secondary focus near the paraboloid vertex $V_\text{p}$. The primary focus is near the sub-reflector and can be seen in \cref{fig:actuators_sub-reflector}.}
      \label{fig:effelsberg_geometry}
    \end{figure*}

    \Cref{fig:effelsberg_geometry} shows the same parameters in \cref{tab:effelberg_geometry} in a scaled diagram of the Gregorian configuration \citep{1961ITAP....9..140H}. A Gregorian telescope has a paraboloid main dish and a ellipsoid secondary dish, where one of the foci of the ellipsoid coincides with the paraboloid focus. Receivers are located in the secondary focus near the paraboloid vertex. The mean surface error\footnote{Technical data  sheet for the 100-m telescope: \url{https://www.mpifr-bonn.mpg.de/245888/specs}.} of the primary reflector is \qty{0.5}{\milli\m} and for the sub-reflector $\leq\qty{60}{\micro\s}$ \citep{2015JAGeo...9....1H}.

    \begin{table}[h]
      \centering
      \caption{Effelsberg telescope geometrical constants.}
      \label{tab:effelberg_geometry}
      \begin{tabular}{lll}
        \hline
        Diameter primary reflector & $D_\text{p}$ & \qty{100}{\m} \\
        Diameter sub-reflector & $D_\text{s}$ & \qty{6.5}{\m} \\
        Effective focal length & $F_\text{eff}$ & \qty{387.394}{\m} \\
        Focal length primary reflector & $ F_\text{p}$ & \qty{29.98}{\m} \\
        Depth primary reflector & $H_\text{p}$& \qty{20.83}{\m}\\
        Semi-major axis sub-reflector &$a$& \qty{14.3050}{\m}\\
        Semi-minor axis sub-reflector & $b$ & \qty{7.3872}{\m}\\
        Eccentricity sub-reflector &$e$ & \num{0.85634} \\
        Support leg length (aperture) &$l$ & \qty{20}{\m} \\
        Support leg width (aperture) &$t$ & \qty{2}{\m} \\
        \hline
      \end{tabular}
    \end{table}

    \section{Gregorian telescope OPD}
    \label{ap:gregorian_telescope_opd}
\begin{figure*}
      \centering
      \includegraphics{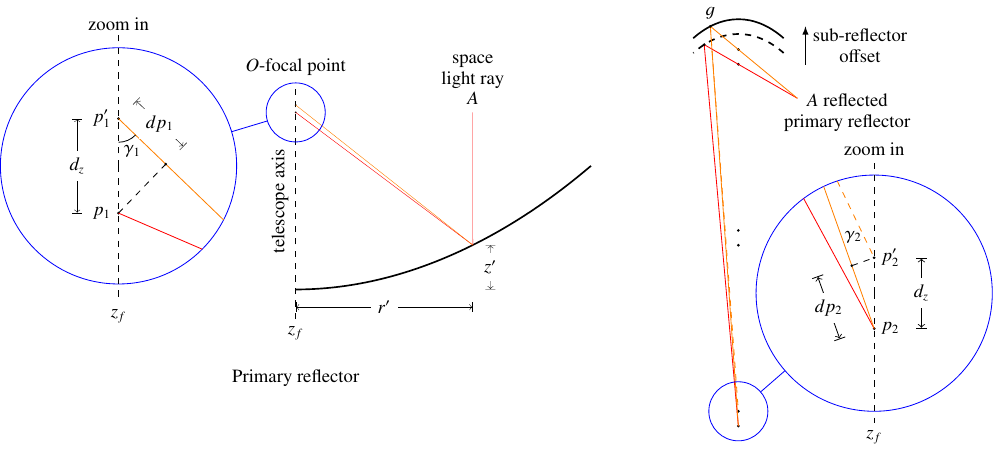}
      \caption{optical path difference (OPD), $\delta(x, y; d_z)$, function of a Gregorian telescope. The red light ray is in-focus and the orange is out-of-focus, given a certain axial offset, $d_z$. Blue lines and circles are zoom-in sections. Top right dashed line represents the sub-reflector offset in the $z_f$-axis. The diagram on the left-hand side corresponds to the primary reflector and its axial defocus. The right-hand side image is the continuation of the same light ray, $A$, reflected on the primary reflector, that reaches the second focus of the ellipse. The axial offset is the same for both foci. The angle $\gamma_2$ exists between the points $p_2$-$g$-$p_2'$.}
      \label{fig:effelsberg_opd}
    \end{figure*}

    The optical path difference (OPD) function for a Gregorian reflector is the same as a Cassegrain reflector, even though their basic geometries are different. 
    To prove this we will rely on \cref{fig:effelsberg_geometry,fig:effelsberg_opd} which show the Gregorian configuration. \Cref{fig:effelsberg_opd} shows the trajectory of a light ray $A$ (far-field plane wave) in the case of zero offset in red ($d_z=\qty{0}{\cm}$) and an added offset in orange ($d_z\neq\qty{0}{\cm}$). The pair of points $p_1, p_2$ and $p_1', p_2'$ are the zero and with offset configurations, respectively. Then the OPD will be given by $dp=dp_1 + dp_2$, and the calculation of each of the terms follows,
    \begin{equation}
      dp_1=d_z\cos\gamma_1=d_z\frac{F_\text{p}-z'}{\sqrt{(F_\text{p}-z')^2+r'^2}}, \label{eq:dp1}
    \end{equation}
    where the second part of \cref{eq:dp1} comes directly from \cref{fig:effelsberg_opd}. The constant $F_\text{p}$ is paraboloid focal length, see \cref{tab:effelberg_geometry}. Such paraboloid cross-section (as in \cref{fig:effelsberg_geometry} left-hand side) can be described in two dimensions as \citep[see][Chapter~7]{stutzman1998antenna},
    \begin{gather}
      \del{\rho'}^2=4F_\text{p}(F_\text{p}-z_f) \quad\text{with}\quad \rho' \leq \frac{D_\text{p}}{2}, \quad\Rightarrow r'^2 = 4 F_\text{p}z' \label{eq:paraboloid} \\ dp_1=d_z\frac{F_\text{p}-z'}{F_\text{p}+z'}=d_z\frac{1-\frac{z'}{F_\text{p}}}{1+\frac{z'}{F_\text{p}}}.
    \end{gather}
    From \cref{fig:effelsberg_opd} we know that distances $p_1$-$p_1'$ and $p_2$-$p_2'$ must be equal, we can then state by analogy,
    \begin{align}
      dp_1 & = d_z\cos\gamma_1, \quad \cos\gamma_1=\frac{1-\mathcal{A}^2}{1+\mathcal{A}^2} \label{eq:defa}\\
      dp_2 & = d_z\cos\gamma_2, \quad \cos\gamma_2=\frac{1-\mathcal{B}^2}{1+\mathcal{B}^2} \label{eq:defb}.
    \end{align}
    Where the following definitions have been used,
    \begin{equation}
      \mathcal{A}\equiv\frac{r'}{2F_\text{p}}=\frac{\sqrt{x^2+y^2}}{2F_\text{p}}\quad\text{and}\quad \mathcal{B}\equiv\frac{r'}{2 F_\text{eff}}=\frac{\sqrt{x^2+y^2}}{2F_\text{eff}},
    \end{equation}
    with $F_\text{eff}$ the effective focal length (equivalent focal length) of the telescope, see \cref{tab:effelberg_geometry}.

    \section{Effelsberg active surface control system}
    \label{ap:effelsberg_active_surface_control_system}

\begin{figure}[t]
      \centering
      \includegraphics{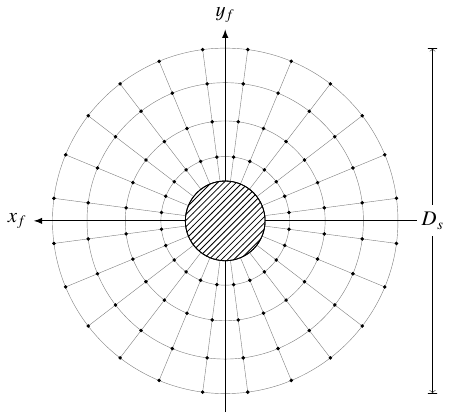}
      \caption{Actuators in the sub-reflector as seen from the  $x_fy_f$-plane. The sub-reflector diameter is labeled as $D_\text{s} = \qty{6.5}{\m}$. Its profile shape is elliptic as seen in \cref{fig:effelsberg_geometry} (left). Dots represent the position of the actuators. The actuators are separated between \qty{15}{\deg} and in four concentric rings: \qtylist{1210;1880;2600;3250}{\mm}. The minimum distance between actuators is \qty{63}{\mm}, and only present in the inner ring. There are $\num{72} + \num{1}$ panels. Diagonal lines in the inner area corresponds to the sub-reflector prime focus receiver blockage (\qty{1.25}{\m} diameter). The out-of-focus holography was only performed with secondary focus receivers.}
      \label{fig:actuators_sub-reflector}
    \end{figure}

    The active surface control system at the Effelsberg telescope is placed in its 6.5-m diameter sub-reflector, with the sub-reflector is the optical conjugate of the main dish. Four concentric rings (\numlist[list-final-separator={, }]{1210;1880;2600} and \qty{3250}{\mm}) are located in its ellipsoid surface (see \cref{fig:actuators_sub-reflector}) where a set of \num{96} actuators deform the surface (only perpendicular displacement over the surface). The active surface contains \num{72} panels and one large circular panel (but elliptic profile) at the center.
    The active surface sub-reflector was installed in 2006 replacing the old sub-reflector with worse surface accuracy \citep{Bach:2007wu}.
    The active surface is able to correct large-scale aberrations by applying perpendicular displacements (to the secondary surface) with the actuators.

\begin{figure*}
      \centering
      \includegraphics{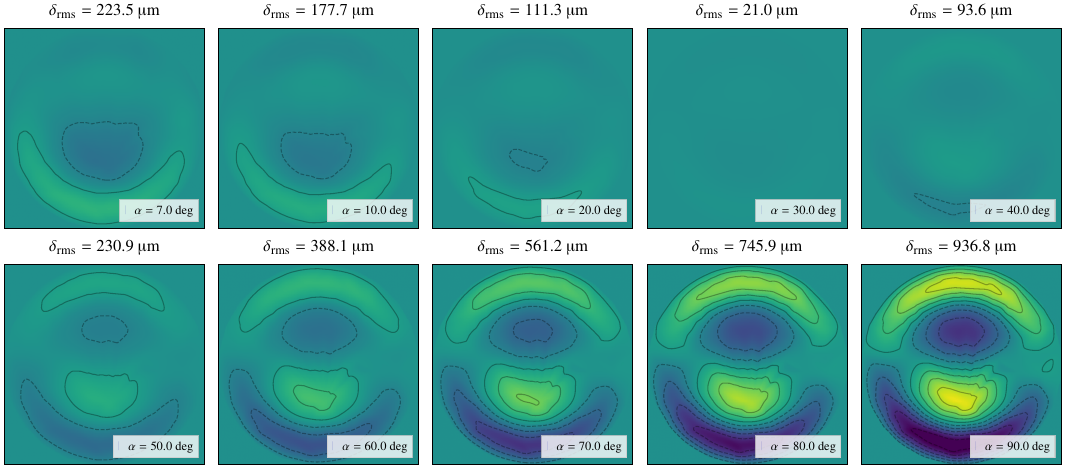}
      \caption{Active surface control system look-up table. There are eleven elevation angles in the look-up table, ${\varphi_\bot}_\text{FEM}$, with the \qty{32}{\deg} set to zero displacements. Each panel shows the elevation angle, $\alpha$, and its rms value in microns, $\delta_\text{rms}$. The contour lines are between \qtylist{-2000;2000}{\micro\m}, with \qty{400}{\micro\m} intervals. The actuators maximum displacement is of $\qty{\pm5}{\mm}$. It is instantly noticeable stronger corrections for an angle closer to \qty{90}{\deg}. The dominant aberrations are first and second coma.}
      \label{fig:fem_lookup}
    \end{figure*}

    The effects of elevation on a parabolic antenna are well known: for each degree in elevation there will be different deformations across its surface \citep{2018ASSL..447.....B}. Such deformations were computed from a FEM model (mechanical structure loads) with eleven elevation angles and stored in a look-up table, \cref{fig:fem_lookup}. From \cref{fig:fem_lookup} it is clear that the main aberration comes from the first and second order coma.

    The look-up table corresponds to \numlist[list-final-separator={, }]{7;10;20;30;32;40;50;60;70;80} and \qty{90}{\deg} (in elevation).
    The \qty{32}{\deg} elevation was taken as a reference point and it is where the actuators have a zero displacement. At this position it is known that the telescope has the closest shape to a perfect paraboloid. The \qty{32}{\deg} elevation reference point was set by performing a phase-coherent holography method, using a geostationary satellite \citep{kesteveen2001effelsberg}.

    The FEM determines loads in all direction of the mechanical structure and then compensate for those with the sub-reflector surface, see \cref{fig:fem_lookup}, most evident at $\alpha=\qty{90}{\deg}$). It is clear that there is a bending moment, given mainly by the reflector's weight, from the top section of the dish that the active surface is trying to correct \citep[see][Chapter~4.5.3]{2018ASSL..447.....B}.
    In general, the FEM model and active surface will only correct for mechanical deformation and deviations from the homologous design of the Effelsberg telescope, and they will not correct for thermal gradients or other source of deformation mechanical deformation.

  \end{appendix}
\end{document}